\documentclass[aps,prd,showpacs,superscriptaddress,twocolumn]{revtex4}
\usepackage{epsfig}
\usepackage[english]{babel}
\usepackage{pgf}
\usepackage{amsfonts,amsmath,amssymb,enumerate,array,amssymb}
\usepackage[latin1]{inputenc}
\usepackage{bm}
\usepackage{fixmath}
\usepackage{amsbsy}
\usepackage{graphicx}
\usepackage{capt-of}
\usepackage{dcolumn}

\usepackage{layouts}
\usepackage{natbib}
\usepackage[hidelinks]{hyperref}
\usepackage{abrev-jour-long}

\def\H{\mathcal{H}}

\def\M{\mathcal{M}}
\def\S{\mathcal{S}}
\def\R{\mathcal{R}}

\begin{document}

\title{Vertical stability of circular orbits in relativistic razor-thin disks}

\author{Ronaldo S. S. Vieira}\email[]{rss.vieira@usp.br}
\affiliation{Instituto de Astronomia, Geof\'isica e Ci\^encias
Atmosf\'ericas, Universidade de S\~ao Paulo, 05508-090,
S\~ao Paulo, SP, Brazil}

\author{Javier Ramos-Caro}\email[]{javier@ufscar.br}
\affiliation{Departamento de F\'isica, Universidade
Federal de S\~{a}o Carlos, S\~ao Carlos, 13565-905 SP, Brazil}

\author{Alberto Saa}\email[]{asaa@ime.unicamp.br}
\affiliation{Departamento de Matem\'atica Aplicada,
Universidade Estadual de Campinas, Campinas, 13083-859  SP, Brazil}


\begin{abstract}
During the last few decades, there has been a growing interest in
  exact solutions of Einstein equations
describing  razor-thin disks. Despite the   progress in the area,
the analytical study of geodesic motion crossing the disk plane in these systems is
not yet so developed. In the present work, we propose a definite vertical stability criterion for circular equatorial
timelike geodesics in static, axially symmetric thin disks,  possibly surrounded by other structures preserving axial  symmetry.
It turns out that the \textit{strong energy condition} for the disk stress-energy content is sufficient for vertical stability of these orbits.
%, as foreseen by several previous preliminary numerical studies.
Moreover, adiabatic invariance of the vertical action variable gives us an approximate third integral of motion for oblique  orbits which deviate slightly
from the equatorial plane. Such new approximate third integral certainly
points to a better understanding of the analytical properties of these orbits.
The results presented here, derived for static spacetimes, may be a starting point to
study the motion around rotating, stationary razor-thin disks. Our results also allow
us to   conjecture that the strong energy condition should be   sufficient to assure transversal stability of periodic orbits for any singular timelike
hypersurface, provided it is invariant under the geodesic flow.
\end{abstract}

%\begin{keywords}
%galaxies: spiral -- galaxies: kinematics and dynamics.
%\end{keywords}
\date{\today}

\maketitle

\section{Introduction}\label{sec:intro}

Many discoidal systems   can be modeled in a first approximation  as axisymmetric structures. Such systems include
disk galaxies (lenticular, barred and spirals) \cite{binneytremaineGD, vanderkruitFreeman2011ARAA} and
accretion disks around black holes and other compact objects
\cite{abramowiczFragileLRR, karasHureSemerak2004CQGra, semerak2002review}.
In all of these systems, the majority  of the orbits of
small objects  are nearly circular and equatorial (e.g.~\cite{sofueRubin2001ARAA}), and hence studying the stability of such orbits is specially relevant. Particularly,
  precise stability criteria for these situations are of great interest.
The study of axially symmetric structures in General Relativity (GR) began in 1966 with the work of Papapetrou \cite{papapetrou1966} on
axially symmetric vacuum solutions. Exact solutions for the Einstein equations
representing thin disks without radial pressure
were first proposed by Morgan and Morgan \cite{morganMorgan1969PR, morganMorgan1970PRD}.
The first applications of the theory of distributions to curved spacetimes
representing razor-thin disks of self-gravitating matter were proposed in \cite{bonnor1968CMaPh, voorhees1972PRD},
together with an initial
study of the stability of circular geodesics \cite{voorhees1972PRD}.
The stability of circular geodesics in smooth axially symmetric stress-energy distributions was first considered
in \cite{bardeen1970ApJ}, in which some remarks about their vertical stability in the limit of an infinitesimally thin disk are presented.
During the last decades,
 many exact solutions representing thin disks were obtained, both in GR
(see for example \cite{bicakLyndenbellKatz1993PRD, gonzalezLetelier1999CQGra, lemosLetelier1994PRD,
letelierOliveira1987JMP, semerakZellerinZacek1999MNRAS, semerakZacekZellerin1999MNRAS, vogtLetelier2003PRD,PhysRevD.86.043008} and the reviews
\cite{karasHureSemerak2004CQGra, semerak2002review})
and in modified theories of gravity
\cite{coimbraLetelier2007PRD,vieiraLetelier2014GRG},
mainly based on   the
formalism of distribution-valued curvature tensors proposed by Taub \cite{taub1980JMP}. For razor-thin disks, such an approach is consistent with the
results about
distributional sources in general relativity obtained by Geroch and Traschen \cite{gerochTraschen1987PRD},
which provide an adequate framework to deal
with distribution-valued curvature tensors with support on spacetime hypersurfaces.

The stability of circular geodesics in smooth axially symmetric spacetimes  is usually analyzed by considering the  geodesic deviation equation (e.g.~\cite{shirokov1973GRG}).  Radial stability can also be analyzed by the relativistic
generalization of Rayleigh's criterion \mbox{\cite{letelier2003PRD, abramowiczKluzniak2005ApSS, vieiraMarekEtal2014PRD}}.
Concerning razor-thin disks, extensive numerical studies
of the dynamics   of timelike geodesic motion were performed in a wide class of
spacetimes, including not only razor-thin disks, but also  rings surrounding black holes
\cite{saaVenegeroles1999PhLA, semerakSukova2010MNRAS, semerakSukova2012MNRAS, sukovaSemerak2013MNRAS, witzanySemerakSukova2015MNRAS}.	Nevertheless,
despite the many existing numerical results,
 the analytical description of motion around these structures is not yet fully developed.

In contrast to the radial stability case, for which essentially the same analysis used for smooth metrics
can be employed,
vertical   stability  cannot be studied in the same framework. The
presence of a $\delta$-like singularity on the plane of the thin disk
\cite{gerochTraschen1987PRD, lemosLetelier1994PRD, gonzalezLetelier1999CQGra, vieiraLetelier2014GRG, taub1980JMP,vogtLetelier2003PRD}
prevents the use of
the geodesic deviation equation and the corresponding first-order perturbation approach, as first pointed out in \cite{semerakZacek2000PASJ}.
 The existing results about vertical stability of circular geodesics in relativistic thin disks
consider only particular cases (dust disk, \cite{voorhees1972PRD}) or are obtained as a limiting case of smooth stress-energy tensors
\cite{bardeen1970ApJ, semerakZacek2000PASJ}.
Although some authors modeled the effects on the geodesic equations due to this $\delta$-like singularity (e.g. \cite{semerakZacek2000PASJ}), the resulting vertical stability criterion is not consistent with a distributional source along the equatorial plane of the spacetime.

In the present work, we solve this apparent contradiction. We present a rigorous framework for the vertical stability analysis of timelike
circular geodesics in static, axially symmetric razor-thin disks, based on
distribution-valued curvature tensors and sources \cite{gerochTraschen1987PRD, taub1980JMP}.
Our framework is conceptually very similar to the Newtonian one
\cite{vieiraRamoscaro2016CeMDA,vieiraRamoscaro2015MG13}, which we now extend to the GR domain.
The paper is organized as follows. Section~\ref{sec:ham-formul}  has
brief revisions on
the Hamiltonian formulation for the geodesic flow in
  axially symmetric, static spacetimes. The rigorous vertical stability conditions are presented
in Section~\ref{sec:vertical-estab}.
In Section~\ref{sec:ThirdIntegral}, we discuss the approximate third integral of motion
obtained from the adiabatic invariance of the vertical action, as well as its prediction
 for the orbit's envelopes.
 All results are compared in Section~\ref{sec:numerical} with numerical experiments for several razor-thin disk models. The conclusions
 and final remarks  are presented in Section~\ref{sec:conclusions}.
 We use natural units (\mbox{$c=G=1$}) and the signature \mbox{$(-,+,+,+)$}.
Greek indices, which vary from 0 to 3, denote spacetime coordinates; Latin indices, varying from
1 to 3 unless otherwise stated, denote only space coordinates.

\section{Hamiltonian formulation for the geodesic flow in static, axially symmetric spacetimes }\label{sec:ham-formul}

Our experience with the Newtonian case
\cite{vieiraRamoscaro2016CeMDA,vieiraRamoscaro2015MG13} leads us to consider the Hamiltonian
formalism for the geodesic flow \cite{ansorg1998JMP,bertschinger,chiconeMashoon2002CQGra} as the basis to
analyze the vertical stability problem. We exploit the fact that the   timelike
geodesic equations admit a Hamiltonian formulation in which the spacetime coordinates ($x^\mu$) can be interpreted as canonical coordinates, with associated
canonical momenta given by \mbox{$p_{\mu}=g_{\mu\nu} \dot x^{\nu}$}. All timelike geodesics are
assumed to be parametrized by their proper time $\tau$, and the upper dots denote derivatives with respect to $\tau$. The Hamiltonian for timelike
geodesics can be written simply as
  \begin{equation}\label{hamiltonian}
   \H (p_\mu, x^\mu)= \frac{1}{2}g^{\mu\nu}p_\mu p_\nu,
  \end{equation}
with the corresponding Hamilton's equations
being fully equivalent to the geodesic equations \cite{ansorg1998JMP, chiconeMashoon2002CQGra}.
The Hamiltonian~(\ref{hamiltonian}) depends on the coordinates $x^{\mu}$ and conjugate momenta $p_{\mu}$,
defining a flow in a 8-dimensional phase space. However, it is possible to reduce the dimensionality of the system
by defining  a new Hamiltonian depending only on  $x^{j}$ and $p_{j}$, i.e. the spatial coordinates and
associated momenta defined by the timelike vector field foliation, which can facilitate the implementation of concepts and techniques used in the Newtonian framework.
Such reduced Hamiltonian is parametrized by the time coordinate $x^0$ and leads to a system of equations equivalent to the geodesics derived from~(\ref{hamiltonian}), but formulated in a 6-dimensional phase space. This can be done by performing the so-called
\textit{isoenergetic reduction} \cite{bertschinger, chiconeMashoon2002CQGra}.

Choosing \mbox{$t = x^0$} as the time coordinate for the reduced Hamiltonian, we can adopt
a parametrization such that \mbox{$dt/d\tau > 0$} for all timelike geodesics.
Solving the constraint  \mbox{$\H = -1/2$} (valid for any timelike geodesic)
for the corresponding canonical momentum $p_0$, one obtains the effective
reduction of   the dimensionality of
the Hamiltonian system governed by (\ref{hamiltonian}).
The   reduced
(and generally time-dependent) Hamiltonian  \mbox{$H=-p_0$} \cite{chiconeMashoon2002CQGra} will have the the form
  \begin{equation}\label{Hred}
   H(t, p_i, x^j) = \frac{g^{0i}p_i}{g^{00}} + \bigg[\frac{1 + \tilde{g}^{ij}p_i p_j}{(-g^{00})}\bigg]^{1/2},
  \end{equation}
where $\tilde g^{ij}$, given by
  \begin{equation}\label{gtilde}
   \tilde{g}^{ij} \equiv g^{ij} - \frac{g^{0i}g^{0j}}{g^{00}},
  \end{equation}
is the inverse of the spatial metric.
It is worth pointing out that, in the Newtonian limit, for which \mbox{$g_{00}=-(1+2\Phi)$},  \mbox{$g^{00}=-(1-2\Phi)$},
\mbox{$g^{i0}=g_{i0}=0$}, \mbox{$\tilde{g}^{ij}=g^{ij}=\delta^{ij}$} and $\Phi$ is the  Newtonian gravitational potential, the
Hamiltonian of Eq.~(\ref{Hred}) reduces to
\begin{equation}
H_{N}=\frac{1}{2}\,p^2 + \Phi + 1.
\end{equation}
Apart from the constant term, this
is the Hamiltonian for Newtonian gravity.

 In the present work,
we are mainly interested in applying the above formalism to
equatorial circular 
orbits in relativistic static, axially symmetric razor-thin disks.
Let us briefly   introduce  some basic properties of the spacetimes corresponding to
such configurations.
We begin by considering the general static, axially symmetric metric in cylindrical coordinates
($t,R,\varphi,z$) \cite{gonzalezLetelier1999CQGra, ujevicLetelier2004PRD}
  \begin{equation}\label{generalmetric}
   ds^2 = - e^\psi dt^2 + e^\gamma(R^2d\varphi^2) + e^\eta(dR^2 + dz^2),
  \end{equation}
where \mbox{$\psi = \psi(R,z)$}, \mbox{$\gamma = \gamma(R,z)$} and \mbox{$\eta = \eta(R,z)$}. Additionally, we assume that the system has reflection symmetry with
respect to \mbox{$z=0$} (equatorial plane), \mbox{$g_{\mu\nu}(R,-z)=g_{\mu\nu}(R,z)$},
along with the conditions ensuring the existence of a razor-thin disk on the equatorial plane:
\mbox{$\partial g_{\mu\nu}/\partial |z|\big|_{z=0}\neq 0$}.
In this case, the isoenergetically reduced Hamiltonian~(\ref{Hred}) can be cast as
  \begin{equation}\label{Hcilindr}
   H = \bigg[e^{\psi-\eta}(p_R^2 + p_z^2) + V_{\rm eff}(R,z)\bigg]^{1/2},
  \end{equation}
where   the effective potential $V_{\rm eff}$ is given by
  \begin{equation}\label{Veffgeral}
   V_{\rm eff} = e^\psi\bigg(1 + \frac{L_z^2}{R^2 e^\gamma}\bigg).
  \end{equation}
The conserved quantity \mbox{$L_z\equiv p_\varphi$} is the  $z$-component of the relativistic angular momentum. The reduced Hamiltonian flow
now evolves in a 4-dimensional phase space
$(p_R,p_z,R,z)$,
  \begin{equation}\label{Hcilindrflow}
   \frac{dp_i}{dt} = - \frac{\partial H}{\partial x^i},\qquad
\frac{dx^i}{dt} = \frac{\partial H}{\partial p_i},\qquad   i=R,z.
  \end{equation}
Timelike circular geodesics correspond to the
fixed point
  \begin{equation}\label{xio}
   \xi_o=(p_R=0,\,p_z=0,\,R=R_o,\,z=0),
  \end{equation}
  for which
  \begin{equation}\label{fpHredgeral}
   \frac{\partial V_{\rm eff}}{\partial R} = 0.
  \end{equation}

Solving Eq.~(\ref{fpHredgeral}) for $L_z$, we obtain the specific angular momentum of a massive test particle moving on an equatorial
circular geodesic with radius $R$
  \begin{equation}\label{LzR}
   L_z(R) = R^2e^{\gamma}\sqrt{\frac{(e^\psi)_{,R}}{e^\psi(R^2 e^\gamma)_{,R} - R^2 e^\gamma(e^\psi)_{,R}}}.
  \end{equation}
This result is in accordance with the expression obtained by Vogt and Letelier \cite{vogtLetelier2003PRD} for isotropic metrics
and with the general expression presented in \cite{letelier2003PRD,vieiraMarekEtal2014PRD}.
%This expression will be crucial for the the study of the vertical stability of the timelike circular equatorial geodesics.

\section{Vertical stability}\label{sec:vertical-estab}

The Hamiltonian~(\ref{Hcilindr}) is a Lyapunov function (an integral of motion, indeed) for the
reduced flow~(\ref{Hcilindrflow}).
Therefore, a sufficient condition for
Lyapunov stability of a fixed point $\xi$ of the flow~(\ref{Hcilindrflow}) is that $\xi$ is a (strict) local minimum of $H$.
For the case of equatorial circular orbits, since \mbox{$e^{\psi-\eta}>0$}, it follows from the expression of $H$ [Eq.~(\ref{Hcilindr})]
that $\xi_o$ given by (\ref{xio}) is Lyapunov stable if the corresponding point in the meridional plane $(R,z)$, $(R_o,0)$, is a
(strict) local minimum of $V_{\rm eff}$ (for a proof of this result in the context of classical mechanics, see \cite{arnoldMmcm}).
Since this result depends only on the continuity of the Hamiltonian (and not on its smoothness, see \cite{arnoldMmcm}), it remains valid
when we introduce a razor-thin disk at the equatorial plane. In this case, the Lyapunov stability condition   is equivalent to
\begin{subequations}\label{stabVeffz}
\begin{equation}\label{stabVeffza}
\frac{\partial V_{\rm eff}}{\partial |z|}(R_o,0) > 0,
\end{equation}
\begin{equation}\label{stabVeffzb}
\frac{\partial^2 V_{\rm eff}}{\partial R^2}(R_o,0) > 0.
\end{equation}
\end{subequations}
We can interpret Eqs.~(\ref{stabVeffza}) and (\ref{stabVeffzb}), respectively, as  vertical
and radial stability conditions. From Eqs.~(\ref{Veffgeral}) and (\ref{LzR}) we obtain
  \begin{equation}\label{d2VdR2geral}
   \frac{\partial^2 V_{\rm eff}}{\partial R^2}\bigg|_{(R_o,0)} = \frac{(e^\psi)_{,R}}{L_z^2}\frac{dL_z^2}{dR},
  \end{equation}
the well-known generalization of Rayleigh's radial stability criterion \cite{letelier2003PRD,vieiraMarekEtal2014PRD}.
In order to avoid repulsive forces, we must consider only regions where \mbox{$(e^\psi)_{,R}>0$} \cite{vieiraMarekEtal2014PRD}.

Let us now focus on the vertical stability of timelike circular geodesics. Consider a razor-thin disk surrounded by an axially symmetric
matter distribution, also symmetric with respect to the equatorial plane.
The vertical stability criterion (\ref{stabVeffz}a) gives us
  \begin{eqnarray}\label{Veffz}
   \frac{\partial V_{\rm eff}}{\partial |z|}\bigg|_{z=0} = e^{\psi}\bigg\{\frac{\partial\psi}{\partial |z|} +
  \frac{L_z^2}{R^2e^\gamma}\Big(\frac{\partial\psi}{\partial |z|} - \frac{\partial\gamma}{\partial |z|}\Big)\bigg\}>0,
  \end{eqnarray}
with the right-hand side evaluated at \mbox{$z=0$}.
By our experience with the Newtonian case \cite{vieiraRamoscaro2016CeMDA, vieiraRamoscaro2015MG13}, we would expect that the above expression could
be related to the matter-energy distribution of the disk.
In fact, we show below that the vertical stability criterion (\ref{Veffz}) is related to the strong energy condition of the disk fluid. Before that,
let us briefly review  the basic features of the stress-energy tensor associated to the razor-thin disk.

\subsection{Stress-energy tensor of a razor-thin disk}

The stress-energy tensor of a system containing a razor-thin disk is given by
  \begin{equation}\label{Tmunu}
   T^\mu_{\ \nu} = Q^\mu_{\ \nu}\hat\delta(z) + D^\mu_{\ \nu},
  \end{equation}
where $Q^\mu_{\ \nu}\hat\delta(z)$ corresponds to the razor-thin disk and $D^\mu_{\ \nu}$ to the smooth matter distribution surrounding it. Also, $\hat\delta(z)$
is the Dirac delta distribution in curved spacetimes, given by (see \cite{vieiraLetelier2014GRG,taub1980JMP})
  \mbox{$\hat\delta(z) \equiv \delta(z)/\sqrt{g_{zz}}$}, where $\delta(z)$ is the usual Dirac delta distribution (in a flat spacetime), which
is related with the Heaviside ``step function'', $\theta(z)$, by the known expression \mbox{$\delta(z)=\theta_{,z}$}. One has \cite{taub1980JMP}
\begin{equation}
  \sqrt{g_{zz}}\, \hat\delta(z)= \theta_{,z} , \qquad \theta(z) = \begin{cases}
                1, & z>0, \\
		1/2, & z=0, \\
		0, & z<0.
               \end{cases}
  \end{equation}
Since $\theta$ does not depend on the spacetime metric, we can write, in this
 adapted coordinate system,
  \begin{equation}
   \theta_{,\mu}(z) = \sqrt{g_{zz}}\,\delta_\mu^z\,\hat\delta(z),
  \end{equation}
where $\delta_\mu^z$ stands for the normal vector  $n_\mu$, associated with the razor-thin disk:
\mbox{$
   n_\mu \equiv \partial z/\partial x^\mu = \delta_\mu^z.
$}
Since  \mbox{$n^\alpha n_\alpha = g^{zz}$}, we have
  \begin{equation}\label{thetamu}
   \theta_{,\mu} = \frac{1}{\sqrt{n^\alpha n_\alpha}}\,n_\mu\hat\delta,
  \end{equation}
  indeed
a different expression from that presented in \cite{taub1980JMP}. We will show now that the above expression for $\theta_{,\mu}$ is in fact the correct
one, which leads to a modification of the stress-energy tensor of the disk, in comparison to the ones considered in
\cite{taub1980JMP,lemosLetelier1994PRD,gonzalezLetelier1999CQGra,
vogtLetelier2003PRD,vieiraLetelier2014GRG}.
First, let us note that $\theta$ does not depend on the particular choice $f(z)$ (instead of $z$) used to describe the thin disk. If $f(z)$ is an increasing function of $z$ with \mbox{$f(0)=0$}, then the corresponding Heaviside function $\theta_f$, Eq.~(\ref{eq:Heavisidephi}), satisfies \mbox{$\theta_f = \theta$}.
Therefore, $\theta_{,\mu}$ cannot depend on the choice of $f$.
The normal vector associated with the function $f$ is
\mbox{  $
   \tilde n_\mu = \partial f/\partial x^\mu
 $}
\cite{taub1980JMP}, with norm \mbox{$\tilde n^\alpha \tilde n_\alpha = \big(f'(z)\big)^2 n^\alpha n_\alpha$}. It follows that
$
   {\sqrt{ n^\alpha  n_\alpha}\tilde n_\mu =  }{\sqrt{\tilde n^\alpha \tilde n_\alpha}}n_\mu,
$
  and from the definition of the delta distribution, $\hat\delta$ does not depend on $f$, so we have that neither does $\theta_{,\mu}$, as expected.

The argument used here for razor-thin disks can be extended to an arbitrary timelike singular hypersurface in a curved spacetime, and the result is
that the correct expression for $\theta_{,\mu}$ is still given by Eq.~(\ref{thetamu}) (see Appendix~\ref{AppendixA}).
This correction leads to an stress-energy tensor
of the form [see Eq.~(\ref{Tmunu})]
  \begin{eqnarray}\label{eq:QmunuGeneral}
   Q_\mu^{\ \nu} &=& \frac{1}{16\pi}\frac{1}{\big(n^\lambda n_\lambda\big)^{1/2}}
\Big\{b_\mu^{\ \sigma}n_\sigma n^\nu - b_\sigma^{\ \sigma}n_\mu n^\nu - b_\mu^{\ \nu}n_\sigma n^\sigma +\nonumber\\
& & + b_\sigma^{\ \nu}n_\mu n^\sigma +
\delta_\mu^\nu\big(b_\sigma^{\ \sigma}n_\alpha n^\alpha - b_\alpha^{\ \sigma}n_\sigma n^\alpha\big)\Big\},
  \end{eqnarray}
(compare to \cite{barrabes1989CQGra}, eq.~17, and \cite{taub1980JMP}, eqs.~2.11--2.14), where the coefficients $b_{\mu\nu}$ are given by Eq.~(\ref{eq:bmunuAppendix})
  \begin{equation}
   [g_{\mu\nu,\alpha}] \equiv b_{\mu\nu}n_\alpha,
  \end{equation}
and \mbox{$b^\mu_{\ \nu} = b_{\alpha\nu}g^{\mu\alpha}$}, \mbox{$b^{\mu\nu} = g^{\alpha\nu}b^\mu_{\ \alpha}$} (see Appendix~\ref{AppendixA}).
For the case of a razor-thin disk, with the simplest choice \mbox{$f(z)=z$}, we have \mbox{$n_\mu = \delta_\mu^z$} and
  \begin{equation}\label{bmunu1}
   b_{\mu\nu} = g_{\mu\nu,z}\Big|_{z=0^+} - g_{\mu\nu,z}\Big|_{z=0^-}
   = 2\frac{\partial g_{\mu\nu}}{\partial |z|}\bigg|_{z=0},
  \end{equation}
The stress-energy tensor of the disk is therefore given by
  \begin{eqnarray}\label{Qmunu}
   Q^\mu_{\ \nu} &=& \frac{\sqrt{g_{zz}}}{16\pi}\bigg\{b^{\mu z}\delta^z_\nu - b^{zz}\delta^\mu_\nu + g^{\mu z}b^z_{\ \nu} -\nonumber\\
  & & - g^{zz}b^\mu_{\ \nu} + b^\alpha_{\ \alpha}(g^{zz}\delta^\mu_\nu - g^{\mu z}\delta^z_\nu)\bigg\}.
  \end{eqnarray}
The formalism introduced here elucidates the rather artificial distinction in the literature between the ``true'' and the ``physical'' stress-energy
tensor of a relativistic razor-thin disk
\cite{gonzalezLetelier1999CQGra, lemosLetelier1994PRD, vogtLetelier2003PRD,vieiraLetelier2014GRG}.
We see that the really relevant quantity is $Q^\mu_{\ \nu}$ [Eqs.~(\ref{eq:QmunuGeneral}), (\ref{Qmunu})], which corresponds indeed to the ``physical'' stress-energy tensor
in the references above.

Note that it follows immediately from Eq.~(\ref{Qmunu}) that
  \begin{equation}\label{Qzz}
   Q^z_{\ z} = 0.
  \end{equation}
Also, substituting the components of the metric (\ref{generalmetric}) in (\ref{Qmunu}), we obtain the nonzero components of $Q_\mu^{\ \nu}$:
%  \begin{eqnarray}
%   Q^t_{\ t} &=& \frac{e^{-\eta/2}}{8\pi}\bigg\{\frac{\partial\gamma}{\partial |z|} +
%  \frac{\partial\eta}{\partial |z|}\bigg\}\bigg|_{z=0},\label{Qtt} \\
%   Q^\varphi_{\ \varphi} &=& \frac{e^{-\eta/2}}{8\pi}\bigg\{\frac{\partial\psi}{\partial |z|} +
%  \frac{\partial\eta}{\partial |z|}\bigg\}\bigg|_{z=0}, \label{Qphiphi} \\
%   Q^R_{\ R} &=& \frac{e^{-\eta/2}}{8\pi}\bigg\{\frac{\partial\psi}{\partial |z|} +
%  \frac{\partial\gamma}{\partial |z|}\bigg\}\bigg|_{z=0}. \label{QRR}
%  \end{eqnarray}
\begin{subequations}\label{eq:QttppRR}
\begin{equation}
Q^t_{\ t} = \frac{e^{-\eta/2}}{8\pi}\bigg\{\frac{\partial\gamma}{\partial |z|} +
  \frac{\partial\eta}{\partial |z|}\bigg\}\bigg|_{z=0},\label{Qtt}
\end{equation}
\begin{equation}
   Q^\varphi_{\ \varphi} = \frac{e^{-\eta/2}}{8\pi}\bigg\{\frac{\partial\psi}{\partial |z|} +
  \frac{\partial\eta}{\partial |z|}\bigg\}\bigg|_{z=0}, \label{Qphiphi}
\end{equation}
\begin{equation}
   Q^R_{\ R} = \frac{e^{-\eta/2}}{8\pi}\bigg\{\frac{\partial\psi}{\partial |z|} +
  \frac{\partial\gamma}{\partial |z|}\bigg\}\bigg|_{z=0}. \label{QRR}
\end{equation}
\end{subequations}
From (\ref{eq:QttppRR}), we have
\mbox{ $
   Q^\mu_{\ \nu} = {\rm diag} (-\sigma, P_R, P_\varphi, 0),
  $}
where $\sigma$ is the energy surface density of the disk and $P_R, P_\varphi$ are its principal pressures in the $R$- and $\varphi$-
directions (see \cite{gonzalezLetelier1999CQGra}). Inverting relations (\ref{eq:QttppRR}), we finally obtain
  \begin{equation}\label{psiz}
   \frac{\partial\psi}{\partial |z|}\bigg|_{z=0} = 4\pi e^{\eta/2}( \sigma + P_\varphi+P_R),
  \end{equation}

  \begin{equation}\label{psi-gammaz}
   \frac{\partial\psi}{\partial |z|}\bigg|_{z=0} - \frac{\partial\gamma}{\partial |z|}\bigg|_{z=0} =
  8\pi e^{\eta/2}(\sigma + P_\varphi).
  \end{equation}

\subsection{The criterion for vertical stability}
The necessary and sufficient condition for
vertical stability of an equatorial circular geodesic of radius \mbox{$R=R_{o}$}
is given by Eq.~(\ref{Veffz}). We now write it in terms of the underlying stress-energy tensor. This condition is obtained after substituting Eqs.~(\ref{psiz}) and (\ref{psi-gammaz}) in (\ref{Veffz}), evaluated at ($R_{o},0$):
  \begin{eqnarray}\label{Veffzsigma}
   \frac{\partial V_{\rm eff}}{\partial |z|}\bigg|_{(R_{o},0)} &=& 4\pi e^{\psi+\eta/2}\Big\{\big( \sigma + P_\varphi+P_R\big) + \nonumber\\
  & & + 2\frac{L_z^2}{R^2e^\gamma}\big(\sigma + P_\varphi\big)\Big\}\bigg|_{(R_{o},0)}>0.
  \end{eqnarray}
Therefore, a sufficient condition for vertical stability of a circular geodesic of radius $R_o$ in the equatorial plane is
   \begin{equation}\label{vertstsigma}
    \begin{cases}
    \sigma + P_\varphi+P_R > 0, \\ \\
    \sigma + P_\varphi >0
    \end{cases}
   \end{equation}
at $R=R_o$, which is  the key result of this work:
The strong energy condition
for the disk fluid (see \cite{hawkingEllisLSS, waldGR}) suffices
 to assure vertical stability of all
circular timelike geodesics in a static razor-thin disk surrounded by
a distribution of matter with the same symmetries.
Such a statement, or the more general condition (\ref{Veffz}) or (\ref{Veffzsigma}), formalizes, in the context of the theory of distribution-valued stress-energy tensors \cite{gerochTraschen1987PRD, taub1980JMP}, the
previous results \cite{bardeen1970ApJ, voorhees1972PRD} about the vertical stability of circular geodesics in the presence of a razor-thin disk.

When restricted to some cases of interest,
conditions (\ref{vertstsigma})   adopt a quite simplified form. For example,
in the case of an isotropic metric (\mbox{$\gamma = \eta$}), the disk
has an isotropic pressure, i.e. \mbox{$P_\varphi=P_R\equiv P$} [see Eqs.~(\ref{Qphiphi}) and (\ref{QRR})], and relations
 (\ref{vertstsigma}) reduce to
 $$
 \mbox{Isotropic Metric}\:\:\rightarrow\qquad
   \begin{cases}
   \sigma + 2P > 0, \\ \\
   \sigma + P> 0.
   \end{cases}
 $$
For a Weyl metric (\mbox{$\gamma = -\psi$}) we have \mbox{$P_R=0$} [from Eq.~(\ref{QRR})]. In this case, the vertical stability conditions
(\ref{vertstsigma}) reduce to
  $$
\mbox{Weyl Metric}\:\:\rightarrow\qquad   \sigma + P_\varphi > 0.
  $$
In the case of a dust thin disk (\mbox{$\gamma = \eta= -\psi$}), for which \mbox{$P_R=P_\varphi=0$}, we have
  $$
  \mbox{Dust Thin Disk}\:\:\rightarrow\qquad \sigma > 0.
  $$

We close this section by briefly mentioning that, writing (\ref{Veffzsigma}) as
  \begin{eqnarray}\label{Veffzsigma2}
   \frac{\partial V_{\rm eff}}{\partial |z|}\bigg|_{(R_{o},0)} &=& 4\pi e^{\psi+\eta/2}\Big\{\Big(1+\frac{L_z^2}{R^2e^\gamma}\Big)
  \big( \sigma + P_\varphi+P_R\big) + \nonumber\\
  & & + \frac{L_z^2}{R^2e^\gamma}\big(\sigma + P_\varphi - P_R\big)\Big\}\bigg|_{(R_{o},0)},
  \end{eqnarray}
we can obtain another sufficient condition for vertical stability of the orbits: \mbox{$\sigma + P_\varphi > |P_R|$}. This criterion, however, is
more stringent than (\ref{vertstsigma}), of which is a particular case.

\section{Approximate third integral of motion}
\label{sec:ThirdIntegral}

Following the same procedure applied to the Newtonian case \cite{vieiraRamoscaro2016CeMDA,vieiraRamoscaro2015MG13},
we obtain an approximate third integral of motion $I_3$ and the amplitude  of the envelope $Z(R)$ of a nearly
equatorial orbit, by considering an approximate separable Hamiltonian near the stable circular orbit of radius $R_o$
and exploring the adiabatic invariance of the action variables (under the assumption of slow $R$-variations). We keep fixed the total energy \mbox{$H=E$} and the $z$-component of the angular momentum $L_z$.

\subsection{Motion near a stable circular orbit}

The approximate Hamiltonian near a circular orbit of radius $R_{o}$ is, to first order in $|z|$  and for small momenta, given by
  \begin{equation}\label{H2approx}
   H^2 \approx A(R_{o})\big(p_R^2 + p_z^2\big) + V_{\rm eff}(R,0) + \frac{\partial V_{\rm eff}}{\partial |z|}(R_o,0)|z|,
  \end{equation}
where \mbox{$A(R_{o})\equiv e^{\psi-\eta}(R_o,0) > 0$} and
all   higher-order terms proportional
to $|z|R$, $|z|R^2$ and so on were discarded. Notice that   $H^2$ is separable,
and therefore $H$ is integrable in these coordinates.
In particular, by introducing
  \begin{equation}
   g(R_{o})\equiv \frac{\partial V_{\rm eff}}{\partial |z|}(R_o, 0),
  \end{equation}
then the quantity
  \begin{equation}\label{Hz2}
   H_z^2 \equiv A(R_{o}) p_z^2 + g(R_{o})|z|
  \end{equation}
is conserved for the Hamiltonian~(\ref{H2approx}).
The corresponding generating function for the vertical action-angle coordinates
is given by
 \begin{equation}
   S_z = \frac{1}{\sqrt{A(R_{o})}}\int^z\sqrt{E_z^2 - g(R_{o})|z|}dz,
 \end{equation}
and the corresponding vertical action variable, in terms of the amplitude of motion \mbox{$Z\equiv z_{max}$}, is
%\begin{equation}
%   J_z = \frac{4E_z^{3}}{3\pi \sqrt{A(R_{o})} g_{R_o}},
%\end{equation}
  \begin{equation}
   J_z = \frac{4 g_{R_o}^{1/2}}{3\pi \sqrt{A(R_{o})}}Z^{(3/2)}.
  \end{equation}
Now, from Eq.~(\ref{Veffzsigma}),
  \begin{eqnarray}
   g(R_{o}) &=& 4\pi e^{\psi+\eta/2}\Big\{\big( \sigma + P_\varphi+P_R\big) + \nonumber\\
  & & +2\frac{L_z^2}{R^2e^\gamma}\big(\sigma + P_\varphi\big)\Big\}\bigg|_{(R_{o},0)}\nonumber
  \end{eqnarray}
and defining
  \mbox{$\tilde{g}(R_{o}) \equiv g(R_{o})/A(R_{o})$},
we finally get
  \begin{equation}
   J_z = \frac{4 \sqrt{\tilde g(R_{o})}}{3\pi}Z^{(3/2)},
  \end{equation}
  which is an approximate adiabatic invariant for our system.

\subsection{Adiabatic invariance of $J_z$}

As in the Newtonian case \cite{vieiraRamoscaro2016CeMDA,vieiraRamoscaro2015MG13},
the approximate adiabatic invariance of $J_z$  implies that the shape of nearly equatorial orbits is
determined by the envelope satisfying
  \begin{equation}\label{envelopes}
   \frac{Z(R)}{Z(R')} = \bigg(\frac{\tilde g(R')}{\tilde g(R)}\bigg)^{1/3},
  \end{equation}
where
  \begin{eqnarray}\label{tilde-g-R}
   \tilde g(R) &=& 4\pi e^{3\eta/2}\Big\{\big( \sigma + P_\varphi+P_R\big) + \nonumber\\
  & & +2\frac{L_z^2}{R^2e^\gamma}\big(\sigma + P_\varphi\big)\Big\}\bigg|_{z=0},
  \end{eqnarray}
provided  the period $T_z$ of vertical oscillations is smaller than the period $T_R$ of radial oscillations.
The corresponding third integral of motion (in addition to $H$ and $L_z$) is given at the orbit's envelope by \mbox{$I_3=Z(R)[\tilde g(R)]^{1/3}$},
where $\tilde g$ is given by Eq.~(\ref{tilde-g-R}).
The vertical amplitude $Z(R)$ depends implicitly on the phase-space coordinates, \mbox{$Z=Z(p_R,p_z,R,z)$}. In terms of these coordinates,
$I_3$ can be written as
  \begin{equation}\label{eq:I3-GR-phasespace}
   I_3=\frac{1}{\big[\tilde{g}(R)\big]^{2/3}}\bigg[\,p_z^2+\tilde{g}(R)|z|\,\bigg],
  \end{equation}
an expression similar to the Newtonian case \cite{vieiraRamoscaro2016CeMDA}.	

At this point, it is worth comparing the GR predictions for the shape of the envelopes (\ref{envelopes}) and for the approximate third integral (\ref{eq:I3-GR-phasespace})
with the Newtonian equivalent ones. Typically, GR razor-thin disks present orbits whose shape depends also on its
angular momentum and on its principal pressures, rather than solely on its surface density.
There is also a dependence on the metric components. Therefore, the surrounding halo can affect the shape
of nearly equatorial orbits through the influence on these metric components. These are purely relativistic
effects, a direct consequence of the nonlinearity of Einstein's field equations.
Although the knowledge of only the physical properties of the thin disk is sufficient to guarantee
vertical stability of circular orbits (once the strong energy condition is satisfied), it is not enough to
determine the shape of nearly equatorial orbits: We must have information about the surrounding matter
(or, equivalently, about the metric components).

%\subsection{Nongeometrized units}

%\red{**** TIRAR ESTA SE\c{C}\~AO? ****}

%\red{*******************************}

%In nongeometrized units, the isoenergetically reduced Hamiltonian reads
%  \begin{equation}
%   \frac{H}{c^2} = \frac{g^{0i}}{g^{00}}\frac{p_i}{c} + \bigg[\frac{1 + \tilde{g}^{ij}\frac{p_i p_j}{c^2}}{(-g^{00})}\bigg]^{1/2},
%  \end{equation}
%with corresponding Hamilton's equations
%  \begin{equation}\label{Heqredc2}
%   \frac{dp_i}{dt} = - \frac{\partial H}{\partial x^i},
%   \qquad
%   \frac{dx^i}{dt} = \frac{\partial H}{\partial p_i}.
%  \end{equation}
%
%The Hamiltonian $H$ has dimensions of specific energy and the $p'$s have dimensions of momentum.
%
%For the static and axially symmetric case,
%  \begin{equation}\label{Hcilindrc2}
%   \frac{H}{c} = \bigg[e^{\psi-\eta}(p_R^2 + p_z^2) + V_{\rm eff}(R,z)\bigg]^{1/2},
%  \end{equation}
%with the effective potential
%  \begin{equation}\label{Veffgeralc2}
%   V_{\rm eff} = c^2e^\psi\bigg(1 + \frac{(L_z/c)^2}{R^2 e^\gamma}\bigg).
%  \end{equation}
%$V_{\rm eff}$ also has dimensions of specific energy.
%
%Eq. (\ref{envelopes}) is still valid, but with $\tilde g(R)$ given by
%  \begin{eqnarray}
%   \tilde g(R) &=& 4\pi G e^{3\eta/2}\Big\{\big( \sigma + \frac{P_\varphi}{c^2}+\frac{P_R}{c^2}\big) + \nonumber\\
%  & & +2\frac{(L_z/c)^2}{R^2e^\gamma}\big(\sigma + \frac{P_\varphi}{c^2}\big)\Big\}\bigg|_{z=0}
%  \end{eqnarray}

\section{Numerical experiments}\label{sec:numerical}

We have tested our results against exhaustive numerical experiments.  
In this Section, we present some typical results   for     
relativistic razor-thin disk models. Specifically,
we compare the envelopes of the low-amplitude
orbits with the prediction of the adiabatic approximation [see Eq.~(\ref{envelopes})] and, most importantly,
check the validity of formula (\ref{eq:I3-GR-phasespace}) for the third integral of motion. All cases we considered correspond to spacetimes with an isotropic metric of the form
  \begin{equation}\label{metrica isotropica}
   ds^2 = -\biggl(\frac{1-f}{1+f}\biggr)^{2}dt^2 + (1+f)^{4} (dR^2 + dz^2 + R^2 d\varphi^2),
   \end{equation}
where $f$ is some function of $R$ and $z$, which determines each particular model.
The properties
of a spacetime given by (\ref{metrica isotropica}) are well known. In particular,
according to (\ref{eq:QttppRR}), we have for such spacetimes
\begin{equation}
\sigma = - \frac{1}{\pi(1+f)^3}\frac{\partial f}{\partial|z|}, \ \ \ P=\bigg(\frac{1}{4(1-f)} - \frac{1}{2}\bigg)\sigma,
\end{equation}
where \mbox{$P = P_\varphi = P_R$}. For all metrics with \mbox{$\partial f/\partial|z| < 0$},
vertical stability is assured. Incidentally, all the usual energy conditions for the disk fluid are also verified in this case for the exterior region of (\ref{metrica isotropica}) ($f<1$).

For the sake of comparison with the Newtonian limit \cite{vieiraRamoscaro2016CeMDA}, it is convenient to
consider
the quantity
\begin{equation}\label{deltaE}
    \Delta E^{2}=\frac{E^{2}-E^{2}_{\rm min}}{E^{2}_{\rm min}}
\end{equation}
which measures the difference between the particle's energy and the minimum of the effective potential, $E^2_{\rm min}$. 
Let us now consider four explicit examples.

\subsection{Vogt-Letelier disks \label{sec:VogtLetelier}} 

The first model we consider is a GR extension, due to Vogt and Letelier \cite{vogtLetelier2003PRD}, of the so-called
 Kuzmin disk in the Newtonian context. It consists basically on a razor-thin
disk in the equatorial plane of an external vacuum (Schwarzschild) field.
In this case, one has
  \begin{equation}\label{f:kuzmin}
   f(R,z)= \frac{m}{2\sqrt{R^2 + (|z|+a)^2}}.
  \end{equation}
The disk is defined by the quantities $m$ and $a$, both  positive   parameters,
whereas the orbits are specified by the constants of motion, $E$ and $L_z$, besides  the initial conditions.

\begin{figure*}[ht]
 \includegraphics[scale=0.4]{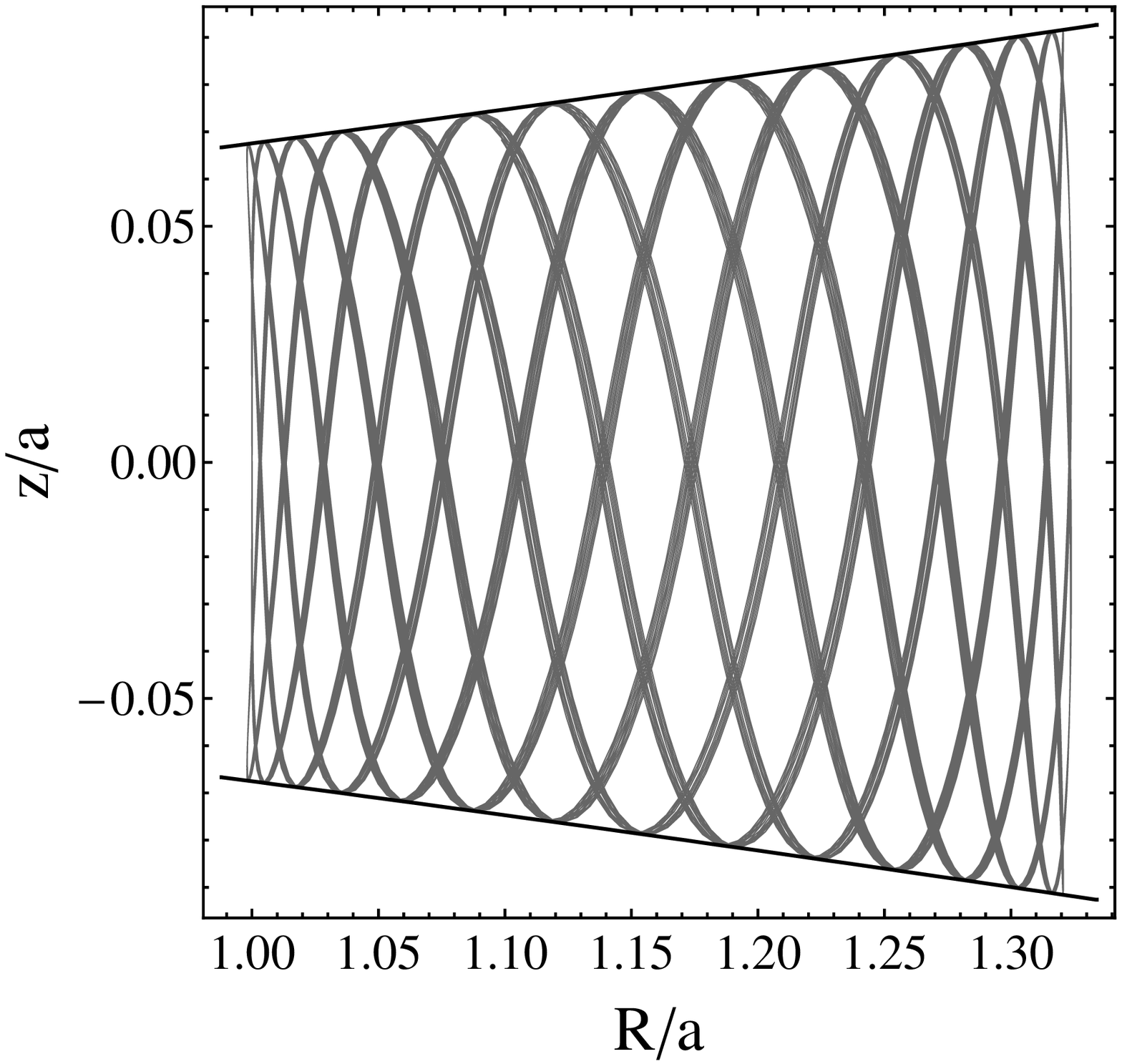}
 \includegraphics[scale=0.4]{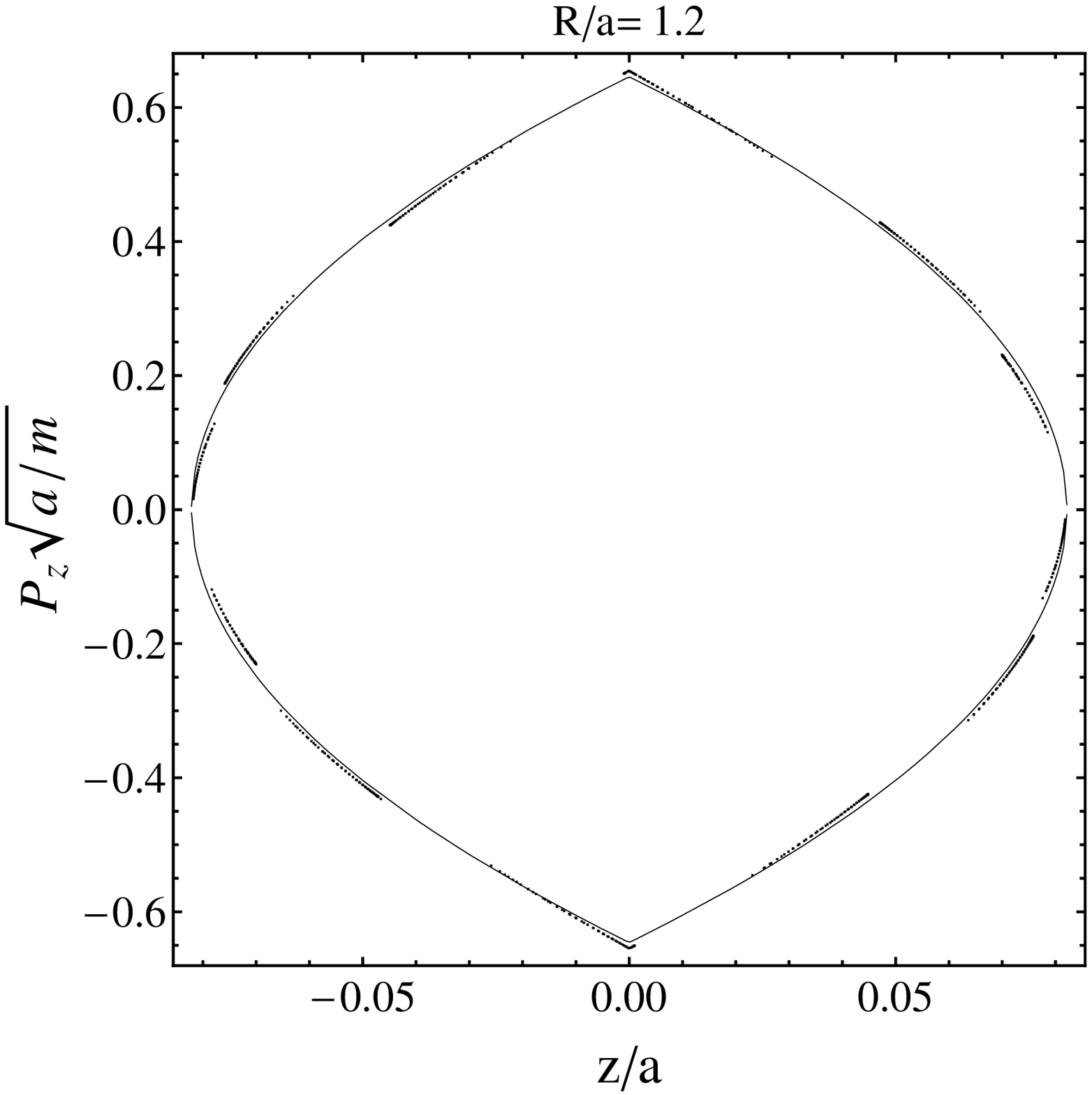}
 \includegraphics[scale=0.7]{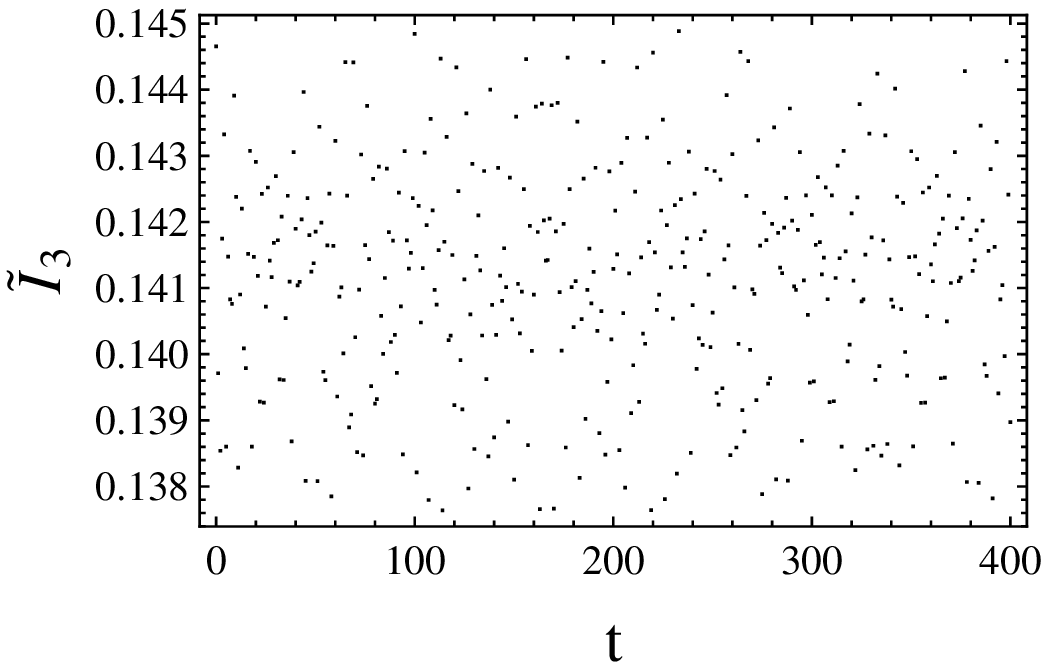}
 \includegraphics[scale=0.6]{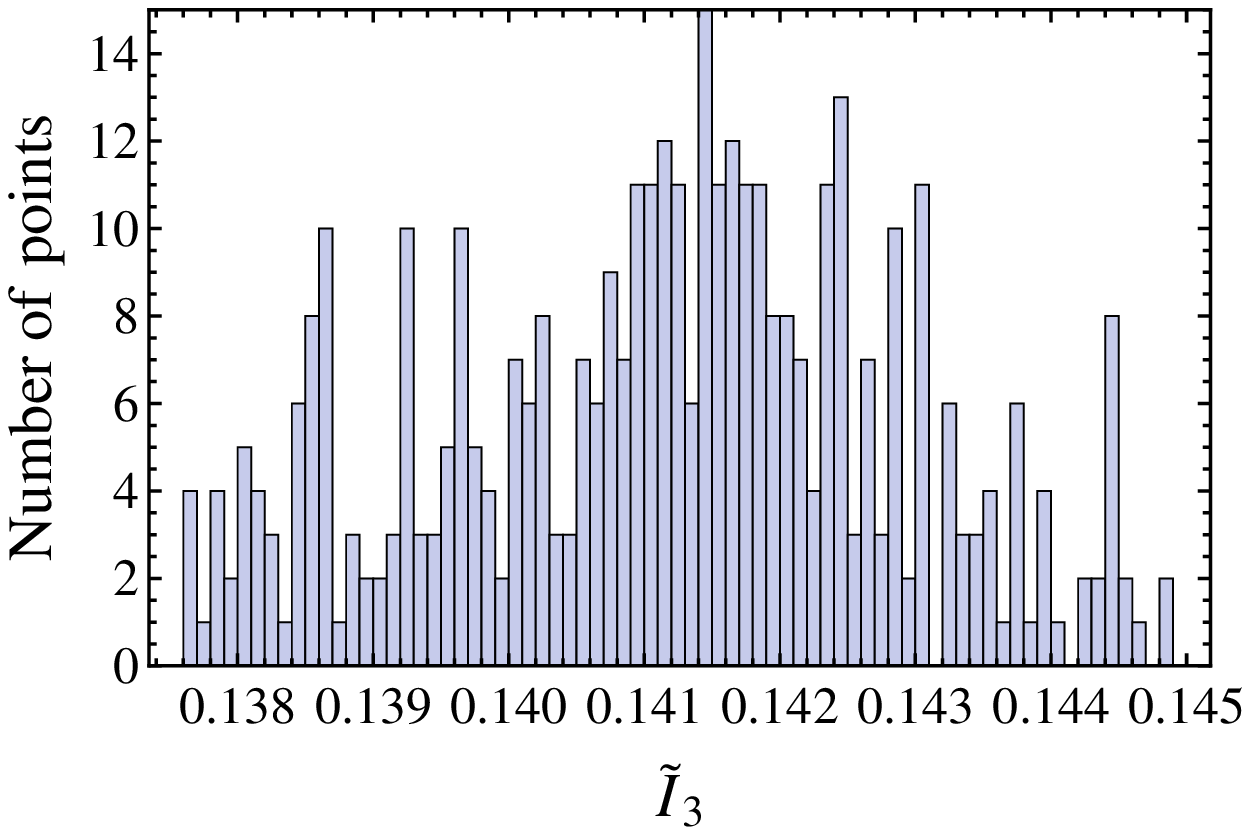}
  \caption{ Top, left: Orbit in  the meridional plane $(R,z)$ for the Vogt-Letelier disk  with \mbox{$m/a=1$}. It has initial conditions \mbox{$R_{o}/a=1$}, \mbox{$P_{R}=0$} and \mbox{$z_{o}/a=10^{-8}$}.
The values of angular momentum and energy are \mbox{$L_z /a=2.2$}, \mbox{$E^{2}=0.6$}, giving \mbox{$\Delta E^2\approx 0.04$}.
The ratio between the vertical and radial average periods is \mbox{$T_z/T_R\approx 0.2$}.
It was calculated from the peaks in the time series of $z(t)$ and $R(t)$.
The predicted envelopes from Eq.~(\ref{envelopes}) are shown by black lines.
In our choice, the predicted and numerically calculated envelopes have the same value
at the upper point of the zero-velocity curve. The predicted envelopes from Eq.~(\ref{envelopes}) are shown by black lines. They agree with the envelopes of the numerically integrated orbits.
Top, right: Poincar\'e section in the surface of constant energy and angular momentum given by the orbit's parameters, with \mbox{$R/a=1.2$}.
The consequents of the orbits are calculated for both \mbox{$P_R>0$} and \mbox{$P_R<0$} and are
given by black dots, whereas the prediction from the average value \mbox{$\tilde{I}_{3,\rm mean}=a^{-2/3}I_{3,\rm mean}=0.141$} 
is given by the solid black line.
We see that Eq.~(\ref{eq:I3-GR-phasespace}) is a good description for the approximate third integral in low-amplitude orbits.
Bottom, left: Time series of \mbox{$\tilde{I}_3(t)=a^{-2/3}I_{3}(t)$}, given by Eq.~(\ref{eq:I3-GR-phasespace}).
It shows small spread along time evolution.
Bottom, right: Histogram for $\tilde{I}_3(t)$.
  }
  \label{fig:GRKuz4}
\end{figure*}

As in the Newtonian case \cite{vieiraRamoscaro2016CeMDA}, where we have obtained for the Kuzmin disk unexpectedly large regions
in which the approximate third integral is valid, we also note that for Vogt-Letelier disks with moderate density profiles
(\mbox{$a/m \gg 1/2$}) the same situation occurs.
Varying the parameters $a$, $E$ and $L_z$, it is possible to investigate whether the aforementioned approximate third
integral is valid along the whole parameter space.
We find that Eq.~(\ref{eq:I3-GR-phasespace}) is always valid along the disk if we consider small enough deviations from
circular motion, however its range of validity depends heavily on the parameters. In particular, for stronger gravity
(\mbox{$a/m\gtrsim 1/2$}), the adiabatic approximation has a rather limited extension near the center of the disk
(low angular momentum). Typically, if the energy of the perturbed orbit is small,
the prediction from (\ref{eq:I3-GR-phasespace}) holds, but once we go to higher values of energy (or higher radial excursions
of the particle) the prediction for the envelopes is poorer.
We present in Fig.~\ref{fig:GRKuz4} a typical disk-crossing orbit oscillating near the equatorial plane and far from
the disk center. In this case
the period of $R$-oscillations is approximately ten times the period of vertical oscillations.
Similar features appear in situations including a halo, but we have to take into account that the volume distribution
can introduce significant deviations from the prediction of Eq.~(\ref{eq:I3-GR-phasespace}). This is the case of the next considered disks.

\subsection{ Kuzmin disk and a Plummer halo \label{sec:KuzPlum}}

\begin{figure*}[ht]
 \includegraphics[scale=0.4]{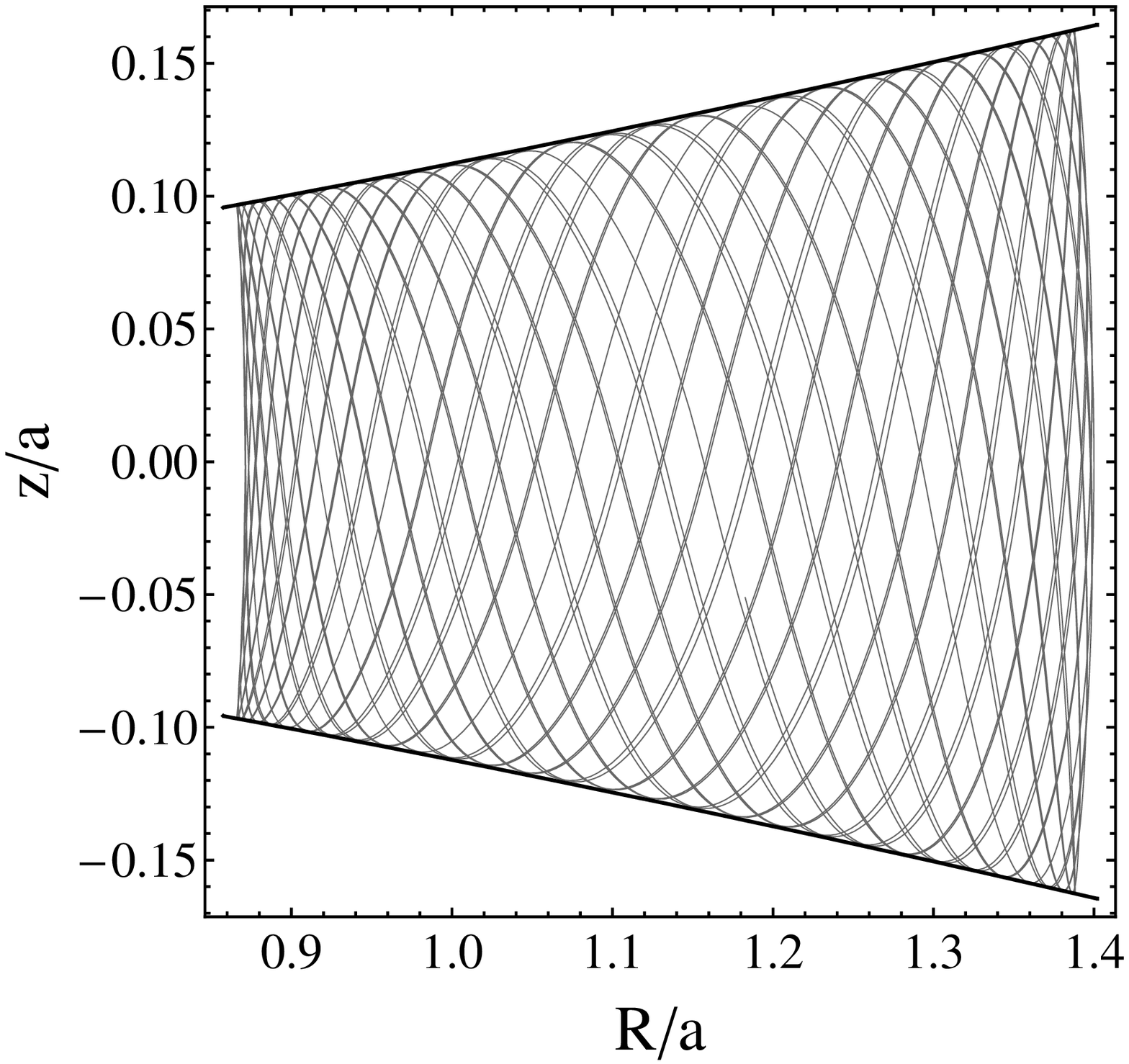}
 \includegraphics[scale=0.4]{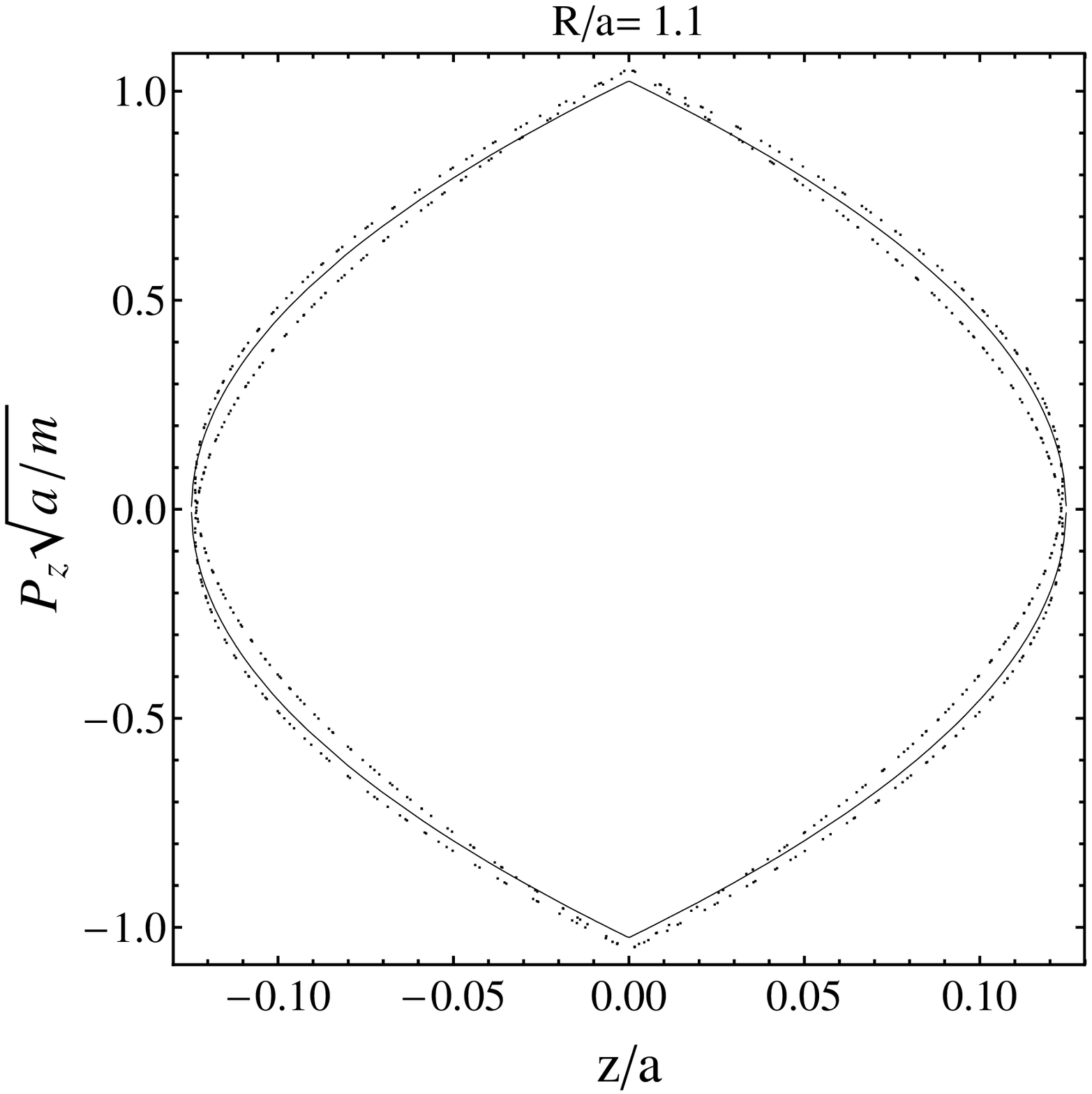}
 \includegraphics[scale=0.7]{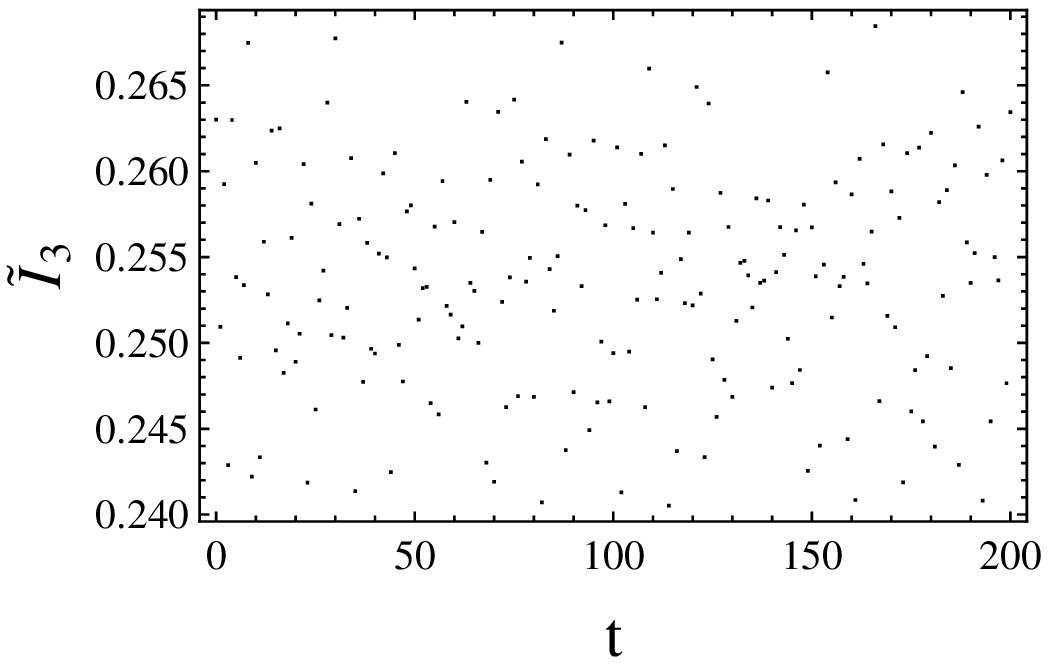}
 \includegraphics[scale=0.6]{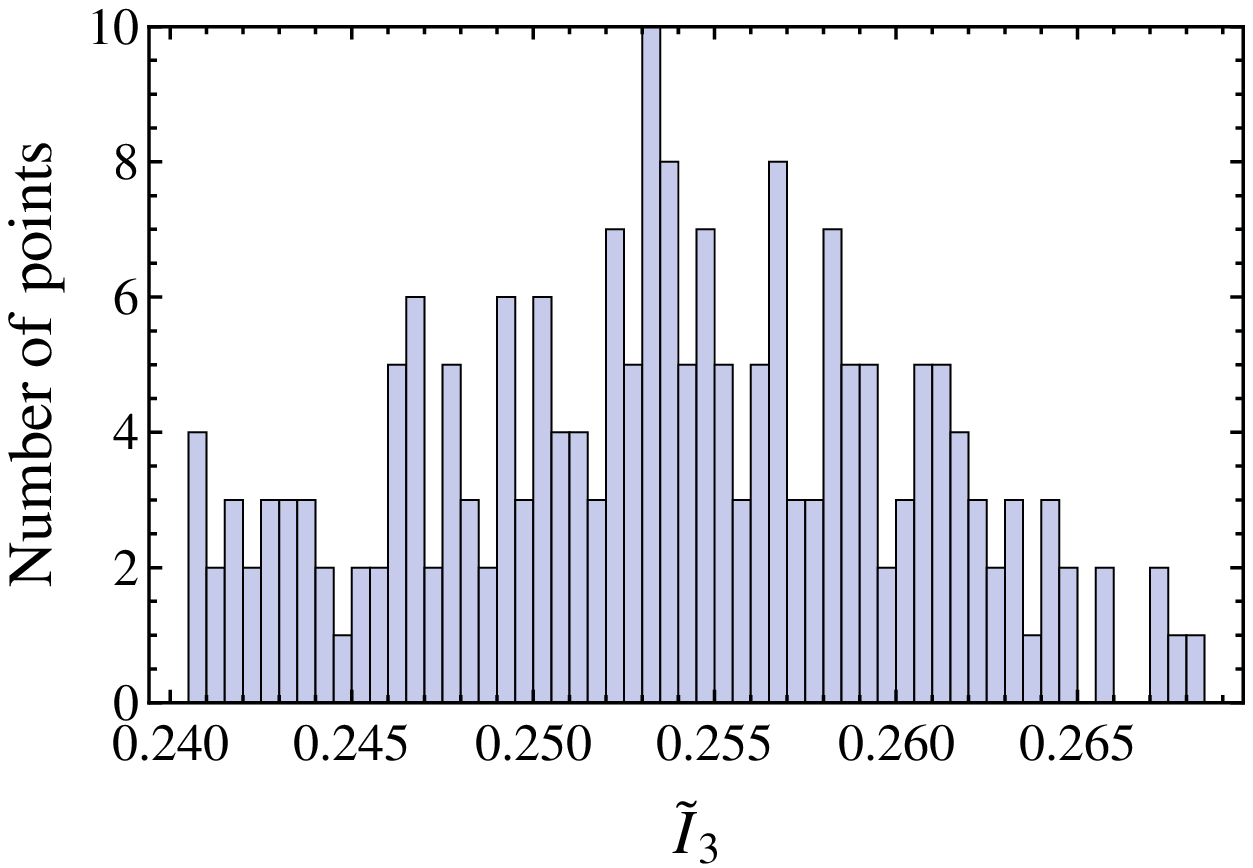}
  \caption{Top, left: Orbit in  the meridional plane $(R,z)$ for the relativistic Kuzmin-Plummer spacetime with \mbox{$m/a=1$},  \mbox{$M/m=0.1$}, and \mbox{$b/a=1$}. 
It has initial conditions \mbox{$R_{o}/a=1.4$}, \mbox{$P_{R}=0$} and \mbox{$z_{o}/a=10^{-8}$}.
The values of angular momentum and energy are \mbox{$L_z /a=2.5$}, \mbox{$E^{2}=0.6$} (\mbox{$\Delta E^2\approx0.08$}).
 For this case, 
\mbox{$T_z/T_R\approx 0.33$}.
The predicted and numerically calculated envelopes are depicted as in Fig.~\ref{fig:GRKuz4}.
Again,  the agreement is  quite satisfactory.
Top, right: Poincar\'e section with \mbox{$R/a=1.1$}, as in Fig.~\ref{fig:GRKuz4}. 
Bottom, left: Time series of \mbox{$\tilde{I}_3(t)=a^{-2/3}I_{3}(t)$}, given by Eq.~(\ref{eq:I3-GR-phasespace}).
It shows small spread along time evolution.
Bottom, right: Histogram for \mbox{$\tilde{I}_3(t)$}. 
The results confirm that Eq.~(\ref{eq:I3-GR-phasespace}) is a good approximate third integral for low-amplitude orbits.
}
  \label{fig:GRKuzPlum7}
\end{figure*}

In spherical coordinates, the Plummer potential is defined as $\phi=-GM[r^{2}+b^{2}]^{-1/2}$ \cite{binneytremaineGD}, so a
general relativistic extension of a Kuzmin disk immersed in a Plummer halo is obtained by
defining
\begin{equation}\label{f:kuzmin-plummer}
   f(R,z)= \frac{m}{2\sqrt{R^2 + (|z|+a)^2}}+\frac{M}{2\sqrt{b^{2}+R^{2}+z^{2}}},
  \end{equation}
where the constants $m$ (disk mass), $M$ (halo mass), $a$ and $b$ are   positive parameters.
Let us call this solution ``relativistic Kuzmin-Plummer spacetime''.
For different combinations of these constants and with the total energy $E$   sufficiently low, 
we found a variety of orbits well described by
Eq.~(\ref{eq:I3-GR-phasespace}), even for the case in which the halo and disk masses are comparable, as in Fig.~\ref{fig:GRKuzPlum7}, where $M/m=10^{-1}$. The calculation details are analogous to those of  Fig.~\ref{fig:GRKuz4}.
For lower ratios $M/m$ we obtain a wider range of validity for relation (\ref{eq:I3-GR-phasespace}). As in the Newtonian case, we find that 
the validity of (\ref{eq:I3-GR-phasespace}) decreases as the halo mass increases, especially in regions where $R$ is of the order of $b$ or smaller.

%KuzPlum7

\subsection{Disk and halo from Buchdahl solution \label{sec:buchdahl}}

\begin{figure*}[ht]
 \includegraphics[scale=0.4]{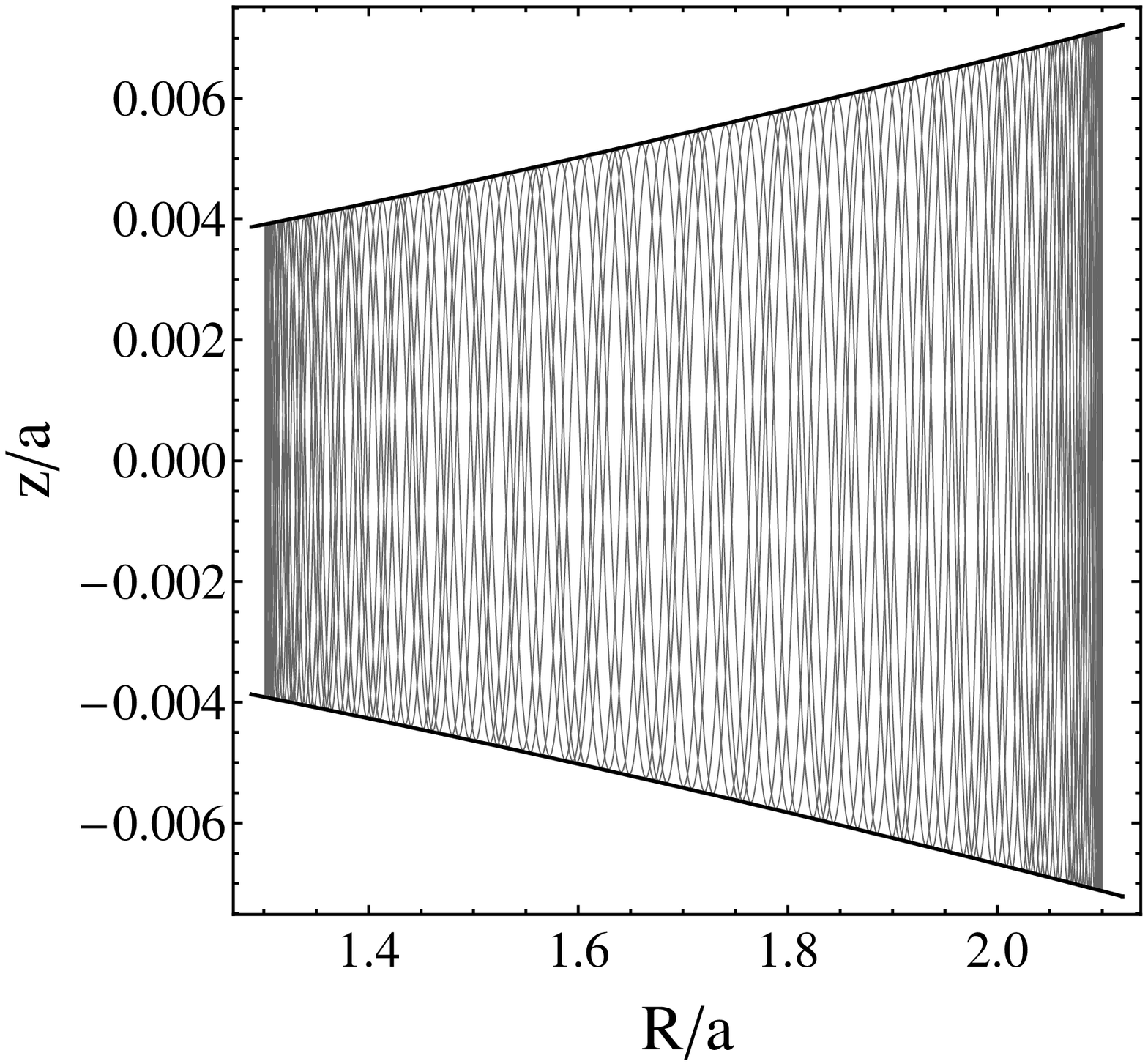}
 \includegraphics[scale=0.4]{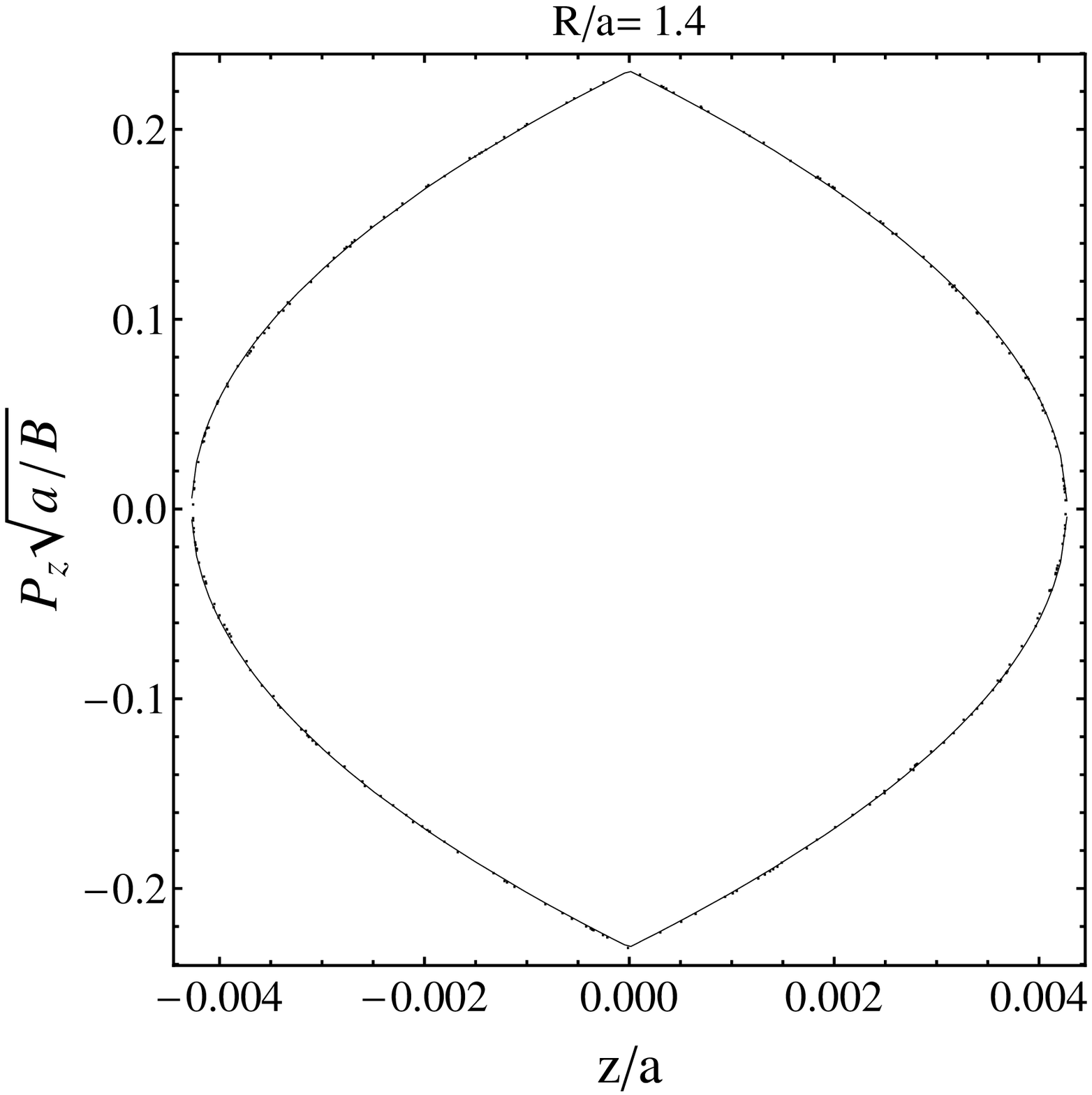}
 \includegraphics[scale=0.7]{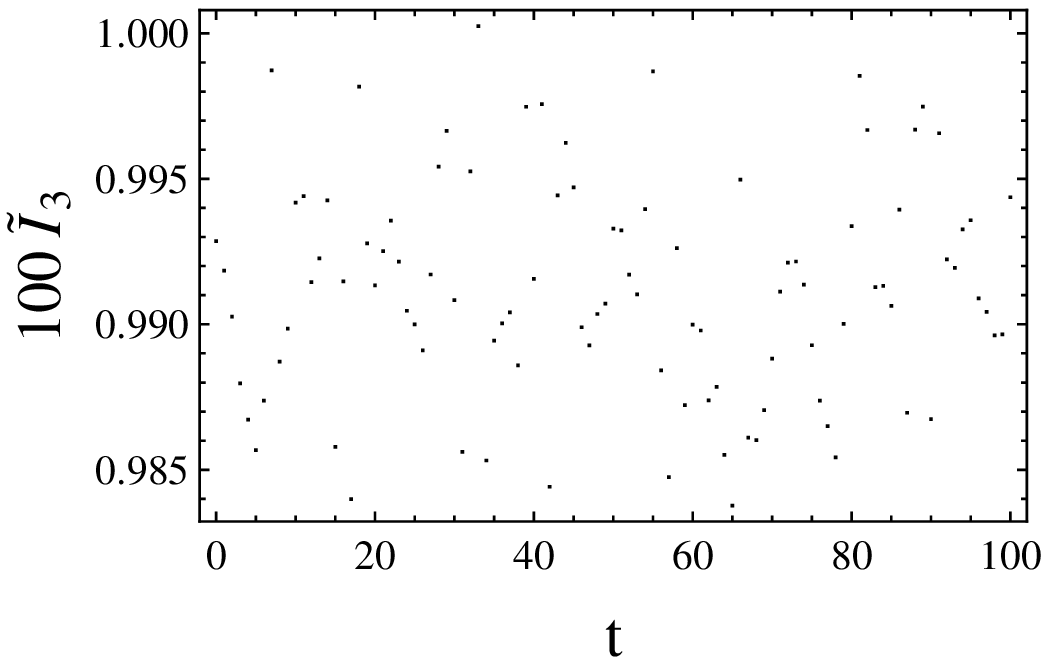}
 \includegraphics[scale=0.6]{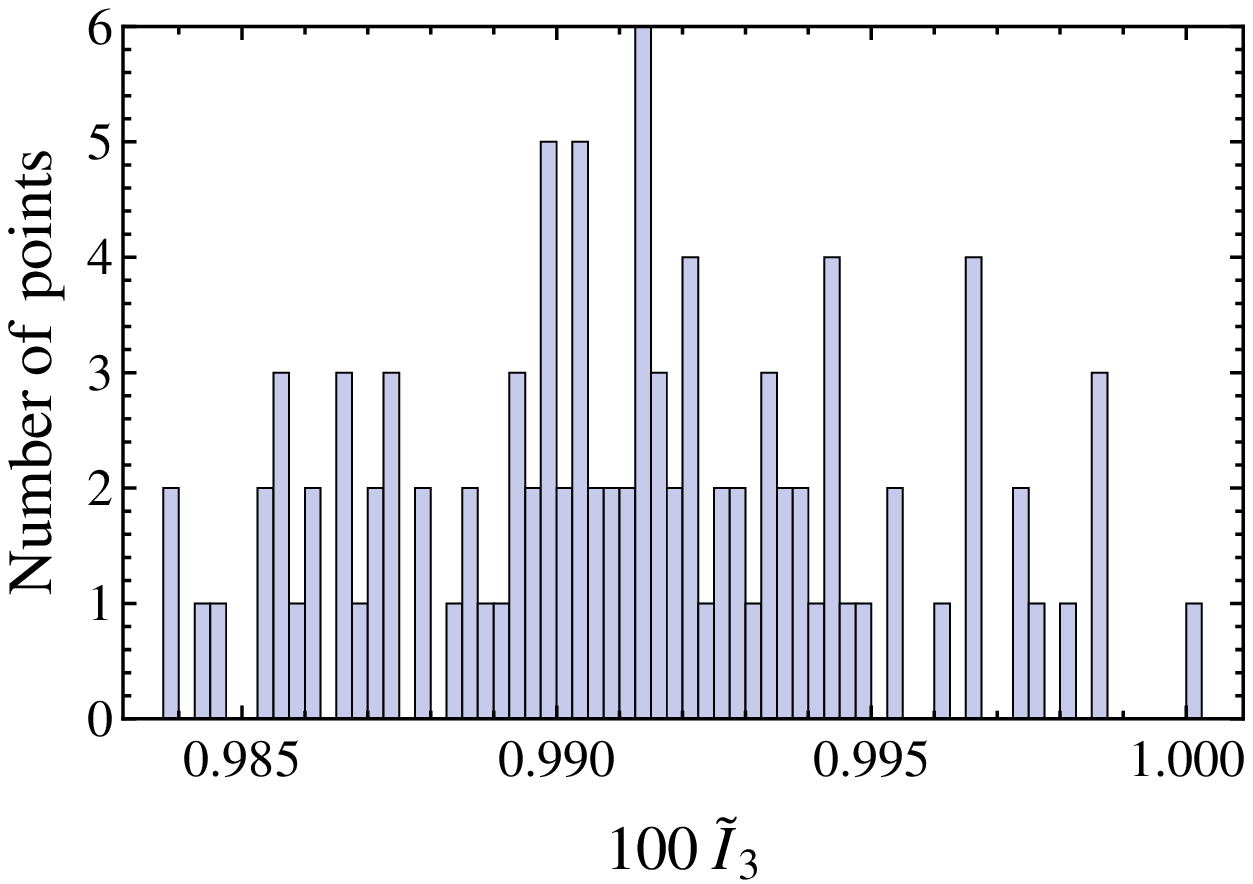}
  \caption{Numerical results for the disk and halo configuration obtained from the Buchdahl
  solution with \mbox{$B=1$} and \mbox{$k/a=2$}. The definitions are the same of Figs.~\ref{fig:GRKuz4} and \ref{fig:GRKuzPlum7}. The orbit initial conditions are  \mbox{$R_{o}/a=2.1$}, \mbox{$P_{R}=0$} and \mbox{$z_{o}/a=10^{-8}$}, with	 \mbox{$L_z /a=5$}, \mbox{$E^{2}=0.92$} (\mbox{$\Delta E^2\approx0.01$}). We have \mbox{$T_z/T_R\approx 0.18$}
The Poincar\'e section is calculated for   with \mbox{$R/a=1.4$} and compared with the predictions
for   \mbox{$\tilde{I}_{3,\rm mean}=0.01$}.
$\tilde{I}_3(t)$ shows small spread along time evolution.
  }
  \label{fig:GRBuchdahl1}
\end{figure*}

Another interesting  example of solution that can be interpreted as an axisymmetric disk with
a surrounding  halo was presented in \cite{vogtLetelier2003PRD}, starting from Buchdahl's solution \cite{1964ApJ...140.1512B}, a   spherically symmetric
model resembling an Emden polytrope of index 5.
It corresponds to
\begin{equation}\label{f:buchdahl}
   f(R,z)= \frac{B}{\sqrt{1+k[R^2 + (|z|+a)^2]}},
  \end{equation}
where $B$, $k$ and $a$ are positive   parameters. We found typically similar results as in the  
Kuzmin-Plummer spacetime.  In particular, the predicted envelopes from (\ref{envelopes})
are verified with good accuracy for low-amplitude orbits, see
  Fig.~\ref{fig:GRBuchdahl1} for further details. 
In this case, all predictions seem to work better
for orbits with small amplitudes obtained from an effective potential whose critical point is away from the center of the configuration.
However, in this situation one can find values of parameters leading to an effective potential characterized by three critical points,
two of them stable and the other one unstable, which is the case of the parameters of Fig.~\ref{fig:GRBuchdahl1}.
In such   conditions,   
we can find significant deviations   between the
predictions and the numerical results, if the energy is large enough.   This case certainly deserves further investigations since
most probably this might be the onset of chaotic behavior in the system.

\subsection{Kuzmin-Hernquist spacetime}

\begin{figure*}[ht]
 \includegraphics[scale=0.4]{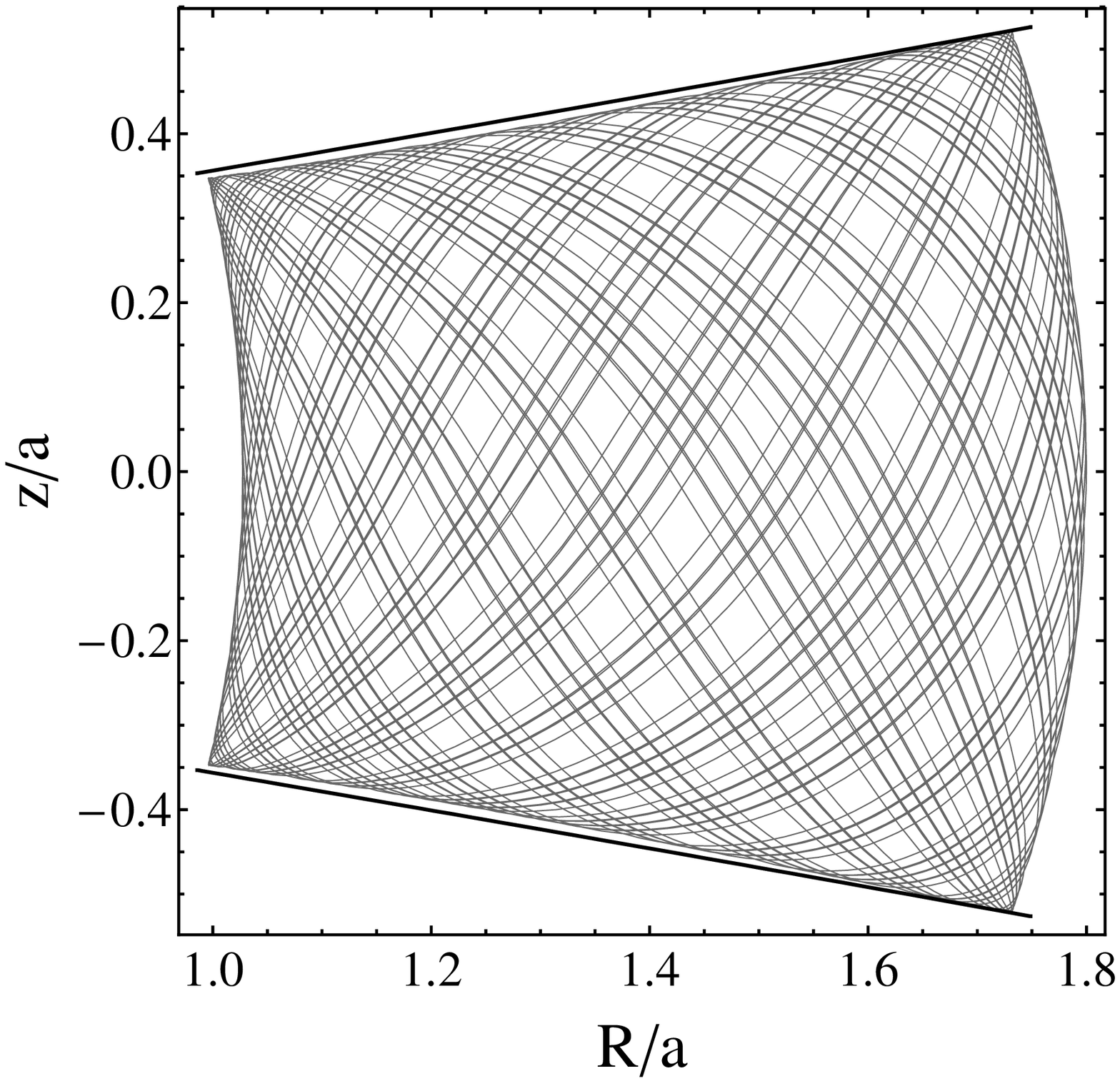}$\:\:\:$
 \includegraphics[scale=0.4]{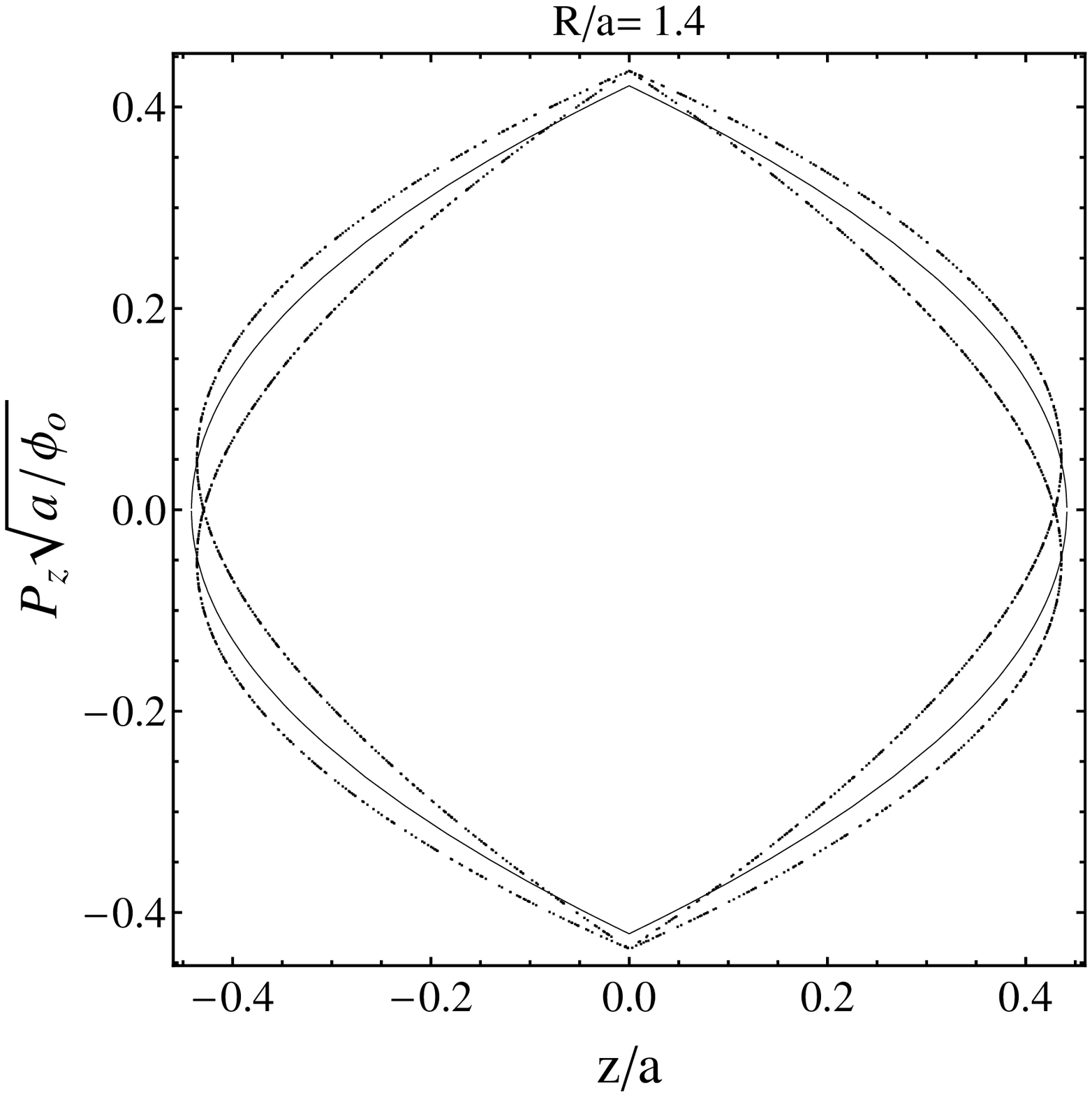}
 \includegraphics[scale=0.7]{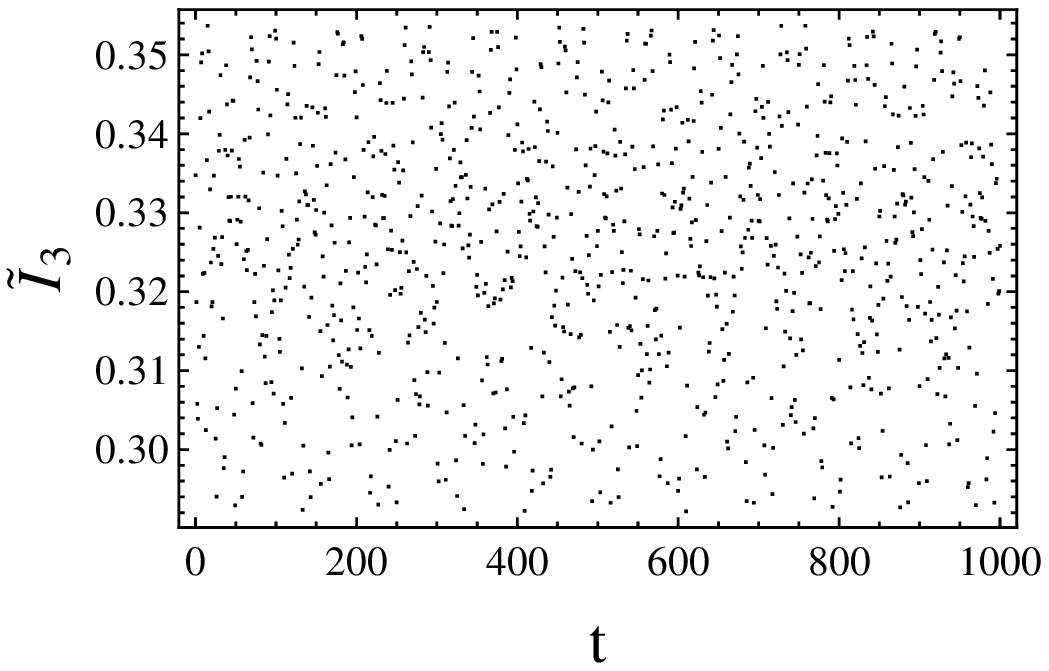}$\:\:\:$
 \includegraphics[scale=0.6]{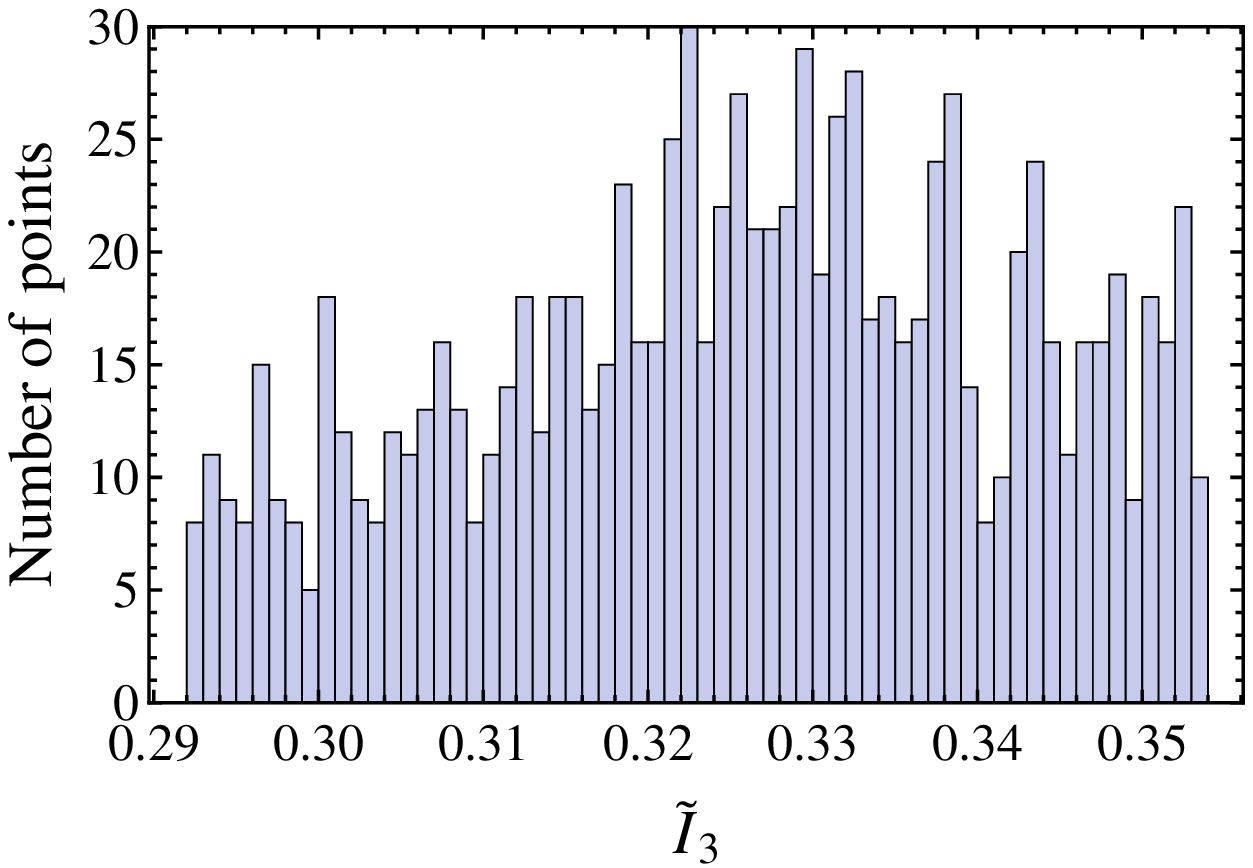}
  \caption{Numerical results for Kuzmin-Hernquist spacetime 
with   \mbox{$\phi_{o}=1$} and \mbox{$b=1$}. The definitions are the same of Figs.~\ref{fig:GRKuz4} and \ref{fig:GRKuzPlum7}. The orbit initial conditions are  \mbox{$R_{o}/a=1.8$}, \mbox{$P_{R}=0$} and \mbox{$z_{o}/a=10^{-8}$}, with   \mbox{$L_z /a=0.74$, $E^{2}=0.6$}, (\mbox{$\Delta E^2\approx0.06$}). We have \mbox{$T_z/T_R\approx 0.7$}.
The Poincar\'e section is calculated for   with \mbox{$R/a=1.4$} and compared with the predictions
for   \mbox{$\tilde{I}_{3,\rm mean}=0.326$}.
$\tilde{I}_3(t)$ shows small spread along time evolution.
  }
  \label{fig:KuzminLike1}
\end{figure*}

Deviations from Eq.~(\ref{eq:I3-GR-phasespace}) caused by halo effects  can also be seen in the 
relativistic extension of the so-called Kuzmin-like potentials \cite{hunter2005NYASA}, i.e
axisymmetric solutions of the form $\phi=\phi(\xi)$, where $\xi=\sqrt{R^{2}+(a+|z|)^{2}}$.
It can be shown that these potentials are produced by the combination of a razor-thin disk at $z=0$ and a volume density
\cite{hunter2005NYASA}.
Here we choose $\phi$ with the form of Hernquist potential, i.e. proportional to $[1+(\xi/b)]^{-1}$, and
the corresponding relativistic extension
is defined by
\begin{equation}\label{f:hernquist}
f(R,z)=\frac{\phi_{o}}{2\left[1+b^{-1}\sqrt{R^{2}+(a+|z|)^{2}}\right]},
\end{equation}
where $\phi_{o}$, $a$ and $b$ are positive   constants.
Let us call this solution Kuzmin-Hernquist spacetime. Its results are depicted in 
Fig.~\ref{fig:KuzminLike1}. Notice that the envelopes (\ref{envelopes}) 
 present in this case a non-negligible difference from the actual vertical amplitudes, 
as well as the predicted surface of section is a bit far from the numerically
obtained one.

It is worth noting that the solutions considered in Sec.~\ref{sec:VogtLetelier} and  Sec.~\ref{sec:buchdahl} are 
also particular cases of the present family.
In all examples, the relativistic extension of the Newtonian gravitational potential $\phi$ was obtained by defining $f=-\phi/2$.
The  Kuzmin disk corresponds to choosing $\phi=-GM/r$.
There is no  halo contribution in this case, and the Newtonian gravitational potential is 
solely due to a razor-thin layer on the equatorial plane.
This feature is inherited by the corresponding relativistic extension defined by (\ref{f:kuzmin}).
On the other hand, Eq.~(\ref{f:buchdahl}) is the result of the choice
$\phi= -2B/\sqrt{1+kr^{2}}$, which corresponds to  
the Emden polytrope of order 4, leading to a  spheroidal volume distribution with a razor thin disk at $z=0$, similar to situation described by  (\ref{f:hernquist}). 
In both cases, we note that as the three-dimensional contribution   grows,
the prediction of Eq.~(\ref{eq:I3-GR-phasespace}) tends to be less accurate, since it only takes into account the fields calculated
at the equatorial plane.

%KuzminLike1-Hernquist

\section{Conclusions}\label{sec:conclusions}

We showed that  the necessary and sufficient condition to assure vertical stability
of equatorial circular geodesics in relativistic razor-thin disks
is \mbox{$\partial V_{\rm eff}/\partial|z|>0$}, evaluated at the disk plane  \mbox{$z=0$}
with \mbox{$L_z=L_z(R_o)$}. It turns out that   the strong energy condition for the singular disk's stress-energy
content suffices to assure the vertical stability of the  equatorial
circular geodesics.
This last result has already been suggested by Bardeen
\cite{bardeen1970ApJ} by considering the thin-limit of a thick
disk,  but without a more rigorous approach based on the
theory of distribution-valued stress-energy tensors as done here.
%We have also shown that the condition
%\mbox{$\sigma + P_\varphi > |P_R|$} is also sufficient   for establishing the vertical stability.

It is worth pointing out that the vertical stability condition obtained here corresponds, in the Newtonian limit, to the expected result \mbox{$\sigma>0$}
\cite{vieiraRamoscaro2016CeMDA,vieiraRamoscaro2015MG13}.
Since the strong energy condition can be considered as one of the possible
relativistic  extensions to the positivity of mass in Newtonian
gravity, we may regard our criterion as a natural counterpart in GR of the Newtonian one.
On the other hand, the approximated third integral
 $I_3$
obtained in Section~\ref{sec:ThirdIntegral} via the adiabatic approximation is not so directly related to the Newtonian result
 \cite{vieiraRamoscaro2016CeMDA,vieiraRamoscaro2015MG13}.
From Eqs.~(\ref{envelopes}) and (\ref{tilde-g-R}) we see that,
although the Newtonian limit gives us the expected result, the nonlinearity of Einstein's field equations
couples the metric functions and the value of the angular momentum of the orbit to the physical components of the thin-disk fluid.
The corresponding third integral  in Newtonian gravity (obtained via the adiabatic approximation) depends only on the surface density of the thin
disk, whereas in GR this nonlinear coupling makes that the corresponding third integral depends implicitly on the properties of the
 surrounding
matter and on the principal pressures of the disk [see Eqs.~(\ref{tilde-g-R})--(\ref{eq:I3-GR-phasespace})].
Therefore, the only case in which $I_3$ is generated solely by the razor-thin disk is
when there is no surrounding matter and, hence, the vacuum field equations are verified outside the singular hypersurface.

Our  definite criterion obtained here solves a quite old apparent paradox in the literature, since different papers have proposed different vertical
stability criteria for circular orbits in the presence of a razor-thin disk \cite{bardeen1970ApJ,voorhees1972PRD,semerakZacek2000PASJ,
zacekSemerak2002CzJP,semerak2002review,karasHureSemerak2004CQGra}. The results of \cite{bardeen1970ApJ,voorhees1972PRD} point in the direction of the
proof presented in this paper, whereas the results of \cite{semerakZacek2000PASJ,zacekSemerak2002CzJP,
semerak2002review,karasHureSemerak2004CQGra}
are mathematically inconsistent and the conclusions derived from them must be reformulated in light of the present approach.
Since the stability criterion obtained and the adiabatic invariance analysis are both
 local, we can extend our formalism to black holes
surrounded by razor-thin disks or rings
\cite{karasHureSemerak2004CQGra, lemosLetelier1994PRD,semerakZacek2000PASJ, semerakSukova2010MNRAS, semerakSukova2012MNRAS},
provided the metric outside
the horizon has the form (\ref{generalmetric}) and the orbit does not cross the horizon. In particular, the results
obtained in \cite{semerakZacek2000PASJ, zacekSemerak2002CzJP}
about vertical instabilities in the inner part of the Lemos-Letelier disk \cite{lemosLetelier1994PRD} (inverted first Morgan and Morgan disk) superposed to a Schwarzschild black hole,
as well as the results in section 4 of the same paper (concerning vertical oscillations of orbits in general razor-thin disks),
must be reformulated in terms of the vertical stability criterion presented here. It turns out that the inverted first Morgan
and Morgan razor-thin disk satisfies the strong energy condition, since this condition
for the Weyl metric [Eq.~(\ref{generalmetric}) with \mbox{$\gamma=-\psi$}] is equivalent to
  \begin{equation}
   \frac{\partial\psi}{\partial |z|}\bigg|_{z=0} > 0
  \end{equation}
by Eq.~(\ref{psiz}), which is satisfied along the whole disk (see eq.~32 of \cite{semerakZacek2000PASJ} or eq.~10 of \cite{zacekSemerak2002CzJP}).
Thus, the conclusions about the radius of the innermost stable circular orbits (ISCOs) in the ``black hole + thin disk''
superposition of \cite{semerakZacek2000PASJ, zacekSemerak2002CzJP} must be
reformulated in terms of the new vertical stability criterion. The authors found a maximum value for the
ratio between the disk and black-hole masses in order to preserve vertical stability of the circular geodesics when approaching the
disk's inner rim. However, as mentioned in their paper \cite{zacekSemerak2002CzJP}, the above result was obtained neglecting the
$\delta(z)$ term.
In view of the new framework introduced here, there is no such limitation. The inner-rim circular orbits are vertically stable
regardless of the mass of the central black hole or of the disk, and therefore the radius of the ISCO in the system under
consideration depends only on the stability under radial perturbations, which was analyzed in \cite{semerakZacek2000PASJ}.
According to
the aforementioned analysis, the inner circular orbits of the disk tend to be more stable with \textit{growing} disk mass,
as obtained in \cite{semerakZacek2000CQGra} for purely equatorial motion,
since the only physical quantity to be considered is the radial epicyclic frequency $\kappa$ (the corresponding radial frequencies are shown
graphically in \cite{semerakZacek2000PASJ}).

According to the results presented here for static razor-thin disks and to the preliminary results of \cite{bardeen1970ApJ} for the
stationary case
(although without a formal proof),
as well as to preliminary computations made by the first author  regarding motion crossing spherical thin shells,
we conjecture that the \textit{strong energy condition} for the stress-energy content of a singular timelike hypersurface is a sufficient condition
to guarantee ``transversal'' stablity of the periodic orbits of this hypersurface, provided that this surface
is invariant under the timelike geodesic flow. This conjecture will be tested in the future for different spacetimes presenting this characteristic.
We also note that, since the reduced Hamiltonian $H$ [Eq.~(\ref{Hcilindr})] is a Lyapunov function for the system (\ref{Hcilindrflow}), all
stability conclusions  are invariant under coordinate transformations
which preserve the time coordinate associated with the timelike Killing vector field.

As a final comment, the extension of Eqs.~(\ref{envelopes})--(\ref{eq:I3-GR-phasespace}) to three-dimensional disks with flattened density and pressure
 profiles is not straightforward. One cannot adopt a ``vertically integrated'' profile (as done in Newtonian gravity; see \cite{vieiraRamoscaro2014ApJ})
  without a careful examination of the term (\ref{tilde-g-R}). The difficulty arises from the nonlinear coupling of the metric with the physical components
  of the disk. Therefore, there is no unique way of extending Eqs.~(\ref{envelopes})--(\ref{eq:I3-GR-phasespace}) to thick disks in GR. A
  more careful analysis is needed in order to compare the possible extensions to three-dimensional structures. This topic will be addressed in future works.

\acknowledgments
The authors thank S\~ao Paulo Research Foundation (FAPESP, grants 2009/16304-3, 2010/00487-9,  2013/09357-9,
and 2015/10577-9)  and CNPq (grant 304378/2014-3) for the financial support.

\appendix

\section{Stress-energy tensor for singular hypersurfaces}
\label{AppendixA}

Let $\M$ be the spacetime manifold and consider a timelike hypersurface $\S$ which splits $\M$ in two parts, i.e.,
\mbox{$\M = \M^+\cup\S\cup\M^-$}.
Let \mbox{$\phi:\M\to\mathbb R$} be a function such that \mbox{$\S = \phi^{-1}(0)$}, with $0$ a regular value of $\phi$.
We choose \mbox{$\M^\pm\equiv \phi^{-1}(\mathbb R^\pm)$}. The ``Heaviside step function'' associated with $\M^+$ is given by \cite{taub1980JMP}
  \begin{equation}\label{eq:Heavisidephi}
   \theta(\phi) = \begin{cases}
                   1, & \phi>0, \\
		   1/2, & \phi=0, \\
		   0, & \phi < 0.
                  \end{cases}
  \end{equation}
Notice that $\theta(\phi)$ does not depend on the particular form of $\phi$, but only on its orientation. The normal vector
associated with $\phi$ is given by \mbox{$n_\mu \equiv \partial\phi/\partial x^\mu$} \cite{barrabes1989CQGra, taub1980JMP},
and the unit normal vector associated with the given orientation is
  \begin{equation}\label{tmu}
   t_\mu \equiv \frac{1}{\sqrt{n^\alpha n_\alpha}}n_\mu.
  \end{equation}
We define the ``jump'' in the derivative of a function $f$ through $\S$ by \cite{taub1980JMP}
  \begin{equation}
   [f]_{,\mu} \equiv f^+_{,\mu} - f^-_{,\mu},
  \end{equation}
and the jump in the metric derivatives by the coefficients $b_{\mu\nu}$:
  \begin{equation}\label{eq:bmunuAppendix}
   [g_{\mu\nu,\alpha}] \equiv b_{\mu\nu}n_\alpha.
  \end{equation}

The Dirac delta distribution $\hat\delta$ with support on $\S$ is defined by \cite{taub1980JMP}
  \begin{equation}\label{inthatdelta}
   \int_\Omega  \hat\delta f \sqrt{-g}d^4x = \int_A f d^3V,
  \end{equation}
where \mbox{$f: \M\to\R$} is an integrable function and \mbox{$A = \Omega \cap \S$}. Here, $d^3V$ stands for the volume element of $S$ induced by the
spacetime metric.
Contrary to what was stated in \cite{taub1980JMP}, the correct formula for the derivative
of the Heaviside function is  \mbox{$\theta_{,\mu} = t_\mu\hat\delta$}, which indeed corresponds
to (\ref{thetamu}).
It is clear from the definition of $\theta$ that $\theta_{,\mu}$ should depend neither on the specific form of $\phi$
(since $\theta$ depends only on $\S$) nor on the form of the spacetime metric (since partial differentiation does not
depend on the existence of a metric). We can readily see that the above equation for $\theta_{,\mu}$ satisfies both requirements.
The corresponding equation given in \cite{taub1980JMP}, however, gives a different result if we choose a different function
$\tilde\phi$ to describe $\S$ or if we consider a different metric $\tilde g$ for the spacetime, maintaining its causal
structure.

We can    prove that the correct formula for $\theta_{,\mu}$ is indeed (\ref{thetamu}).
Given a compact domain
\mbox{$\Omega\subset \M$} such that \mbox{$\Omega\cap\S = A$},  $\theta_{,\mu}t^\mu$ is well defined in $\Omega$ and
  \begin{eqnarray}
  & & \int_\Omega \theta_{,\mu}t^\mu f \sqrt{-g}d^4x
  = \int_\Omega \left[(\theta f \zeta^\mu)_{;\mu} - \theta(f \zeta^\mu)_{;\mu} \right]\sqrt{-g}d^4x \nonumber\\
  & & \quad   = \int_\Omega \left(\theta f \zeta^\mu\right)_{;\mu}\sqrt{-g}d^4x - \int_{\Omega^+} (f \zeta^\mu)_{;\mu}\sqrt{-g}d^4x,
  \end{eqnarray}
where \mbox{$\Omega^+ \equiv \Omega\cap \M^+$} and $\zeta^\mu$ is a smooth spacelike vector field  satisfying \mbox{$\zeta^\mu=t^\mu$}  over $\S$.
We have that
  \begin{equation}
   \int_{\Omega^+} (f \zeta^\mu)_{;\mu}\sqrt{-g}d^4x \xrightarrow{Vol(\Omega^+)\to 0} 0.
  \end{equation}
Moreover,
  \begin{equation}
   \int_\Omega \left(\theta f \zeta^\mu\right)_{;\mu}\sqrt{-g}d^4x = \int_{\partial\Omega^+} f \zeta^\mu d^3 V_\mu,
  \end{equation}
where $d^3 V_\mu$ is the oriented volume in $\partial\Omega$ induced by the metric and \mbox{$\partial\Omega^+ \equiv \partial\Omega\cap\M^+$}.
If $\partial\Omega^+$ ``tends to'' $A$, we have that \mbox{$\zeta^\mu d^3 V_\mu\to d^3V$}, where $d^3V$ is the volume element of $A$.
Therefore, in the ``limit'' \mbox{$\partial\Omega^+\to A$},
  \begin{equation}\label{intthetamu}
   \int_\Omega \theta_{,\mu}t^\mu f \sqrt{-g}d^4x \to \int_A f d^3V.
  \end{equation}
Since $\theta$ is constant in $\Omega^+$, we have that \mbox{$\theta_{,\mu}=0$} in $\Omega^+$. Also, the right-hand side of Eq.~(\ref{intthetamu}) does not depend on the particular form of $\Omega$, being valid for an arbitrary region. These two assertions
lead to the result
  \begin{equation}\label{intthetamuequality}
   \int_\Omega \theta_{,\mu}t^\mu f \sqrt{-g}d^4x = \int_A f d^3V,
  \end{equation}
and thus \mbox{$\theta_{,\mu}t^\mu = \hat\delta$}, where $\hat\delta$ is defined by (\ref{inthatdelta}).
The differential of $\theta$ over $\S$ points to its normal direction $n^\mu$, parallel to $t^\mu$. Thus
\mbox{$
   \theta_{,\mu} = h n_\mu\hat\delta,
$}
where \mbox{$h\in C^\infty(\S)$}. It follows that
\mbox{$
   \theta_{,\mu}t^\mu = h n_\mu t^\mu\hat\delta,
 $}
and by (\ref{tmu}) it follows that
 \mbox{$h = 1/\sqrt{n^\alpha n_\alpha}$}\,, leading finally to   (\ref{thetamu}).

\footnotesize

%\bibliography{stabgr_refs}

%\bibliographystyle{physrev}
%\bibliographystyle{h-physrev4}
%\bibstyle{h-physrev4}
%\bibliographystyle{apsrev}

\end{document}